\documentclass[longauth]{aa}
\usepackage[utf8]{inputenc}
\usepackage{tabularray}
\usepackage[varg]{txfonts}
\usepackage{overpic}
\usepackage{tabularx}
\usepackage{multirow}
\usepackage{verbatim} 
\usepackage{algpseudocode}
\usepackage{cases}
\usepackage[breaklinks, colorlinks, citecolor=blue, linkcolor=blue]{hyperref}
\usepackage[usenames, dvipsnames]{color}
\usepackage{amsmath}
\usepackage{graphicx,textcomp}
\graphicspath{{./fig/}}
\usepackage[english]{babel}
\usepackage{siunitx}
\usepackage{float}
\usepackage{url}
\usepackage{longtable}
\usepackage{fancyhdr}
\usepackage{natbib}
\usepackage[normalem]{ulem}
\usepackage[version=4]{mhchem}
\usepackage{graphicx}
\usepackage{txfonts}
\usepackage{supertabular}
\usepackage{longtable}

\newcommand{\planetb}{TOI-3493\,b}

\newcommand{\ppr}{$R_p$}

\newcommand{\ppteq}{$T_\mathrm{eq}$}

\newcommand{\pbr}{$R_\mathrm{b}$}

\newcommand{\pbteq}{$T_\mathrm{eq}$}

\newcommand{\Rea}{$\mathrm{R_{\oplus}}$}

\newcommand{\pptequ}{K}

\newcommand{\tess}{TESS}
\newcommand{\jwst}{JWST}

\makeatletter

\def\instrefs#1{{\def\scsep{\def\scsep{,}}\@for\w:=#1\do{\scsep\ref{inst:\w}}}}

\renewcommand{\inst}[1]{\unskip$^{\instrefs{#1}}$}

\renewcommand*\aa@pageof{, page \thepage{} of \pageref*{LastPage}}

\makeatother

\begin{document}

\title{TOI-3493\,b: A planet with a Neptune-like density transiting a bright G0-type star}
\titlerunning{A planet with a Neptune-like density TOI-3493~b}
\subtitle{}
\author{
P. Chaturvedi \inst{tls,tifr}
\and E. Goffo \inst{tls,tori} 
\and D. Gandolfi \inst{tori} 
\and C.M. Persson \inst{os}
\and A.P. Hatzes \inst{tls}
\and G. Nowak \inst{ncu} 
\and A.~Bonfanti \inst{graz}
\and A.~Bieryla \inst{cfa}
\and W.D. Cochran\inst{ut,cps}
\and K.~A. Collins \inst{cfa} 
\and S.B.~Fajardo-Acosta \inst{caltech}
\and S.B.~Howell \inst{nasa}
\and J.M.~Jenkins \inst{nasa} 
\and J.~Korth \inst{lund}
\and J. Livingston \inst{abc,naoj,asp}
\and E. Pall\'e \inst{iac,ull}
\and S.N.~Quinn \inst{cfa} 
\and R.~P. Schwarz \inst{cfa} 
\and S.~Seager \inst{mit,mit-earth,mit-aero} 
\and A. Shporer \inst{mit}
\and K.G.~Stassun \inst{vu}
\and S. Striegel \inst{seti} 
\and V. Van Eylen \inst{ucl}
\and C.N.~Watkins \inst{cfa}
\and J.N.~Winn \inst{princ} 
\and C.~Ziegler \inst{asu}
}
\institute{
\label{inst:tls}Th\"uringer Landessternwarte Tautenburg, Sternwarte 5, 07778 Tautenburg, Germany \\
\email{priyanka.chaturvedi@tifr.res.in}
\and
\label{inst:tifr}Department of Astronomy and Astrophysics, Tata Institute of Fundamental Research, Mumbai, India, 400005
\and
\label{inst:tori}Dipartimento di Fisica, Universit\`a degli Studi di Torino, via Pietro Giuria 1, I-10125, Torino, Italy
\and
\label{inst:os}Department of Space, Earth and Environment, Chalmers University of Technology, Onsala Space Observatory, 43992 Onsala, Sweden
\and
\label{inst:ncu}Institute of Astronomy, Faculty of Physics, Astronomy and Informatics, Nicolaus Copernicus University, Grudzi\c{a}dzka 5, 87-100 Toru\'n, Poland
\and
\label{inst:graz}Space Research Institute, Austrian Academy of Sciences, Schmiedlstrasse 6, A-8042 Graz, Austria
\and
\label{inst:cfa}Center for Astrophysics \textbar \ Harvard $\&$
Smithsonian, 60 Garden Street, Cambridge, MA 02138, USA
\and
\label{inst:ut}McDonald Observatory, The University of Texas, Austin Texas USA
\and
\label{inst:cps}Center for Planetary Systems Habitability, The University of Texas, Austin Texas USA
\and
\label{inst:caltech}Caltech/IPAC, Mail Code 100-22, Pasadena, CA 91125, USA
\and
\label{inst:nasa}NASA Ames Research Center, Moffett Field, CA 94035, USA
\and
\label{inst:lund}Lund Observatory, Division of Astrophysics, Department of Physics, Lund University, Box 118, 22100 Lund, Sweden
\and
\label{inst:abc}Astrobiology Center, 2-21-1 Osawa, Mitaka, Tokyo 181-8588, Japan
\and
\label{inst:naoj}National Astronomical Observatory of Japan, 2-21-1 Osawa, Mitaka, Tokyo 181-8588, Japan
\and
\label{inst:asp}Astronomical Science Program, Graduate University for Advanced Studies, SOKENDAI, 2-21-1, Osawa, Mitaka, Tokyo, 181-8588, Japan
\and
\label{inst:iac}Instituto de Astrof\'isica de Canarias (IAC), Calle V\'ia L\'actea s/n, 38205 La Laguna, Tenerife, Spain
\and
\label{inst:ull}Departamento de Astrof\'isica, Universidad de La Laguna (ULL), 38206 La Laguna, Tenerife, Spain
\and
\label{inst:mit}Department of Physics and Kavli Institute for Astrophysics and Space Research, Massachusetts Institute of Technology, Cambridge, MA 02139, USA
\and
\label{inst:mit-earth}Department of Earth, Atmospheric and Planetary Sciences, Massachusetts Institute of Technology, Cambridge, MA 02139, USA
\and
\label{inst:mit-aero}Department of Aeronautics and Astronautics, MIT, 77 Massachusetts Avenue, Cambridge, MA 02139, USA
\and
\label{inst:vu}Department of Physics and Astronomy, Vanderbilt University, Nashville, TN 37235, USA
\and
\label{inst:seti}SETI Institute, Mountain View, CA 94043 USA/NASA Ames Research Center, Moffett Field, CA 94035 USA
\and
\label{inst:ucl}Mullard Space Science Laboratory, University College London, Holmbury St Mary, Dorking, Surrey RH5 6NT, UK
\and
\label{inst:princ}Department of Astrophysical Sciences, Princeton University, Princeton, NJ 08544, USA
\and
\label{inst:asu}Department of Physics, Engineering and Astronomy, Stephen F. Austin State University, 1936 North St, Nacogdoches, TX 75962, USA
}
\date{Received December 2024 / Accepted dd Month 2025}
\abstract{
We report the discovery of TOI-3493 b, a sub-Neptune-sized
planet on an $8.15$-d orbit transiting the bright (V=9.3) G0 star HD 119355 (aka TIC 203377303) initially identified by NASA's \tess ~space mission. With the aim of confirming the planetary nature of the transit signal detected by \tess\  and determining the mass of the planet, we performed an intensive Doppler campaign with the HARPS spectrograph, collecting radial velocity measurements. 
We found that TOI-3493~b lies in a nearly circular orbit and has a mass of $9.0\pm{1.2}~M_\oplus$ and a radius of $3.22\pm{0.08}~R_\oplus$, implying a bulk density of $1.47^{+0.23}_{-0.22}~\mathrm{g~cm^{-3}}$,
consistent with a composition comprising a small solid core surrounded by a
thick H/He-dominated atmosphere.
}
\maketitle
\section{Introduction}
\label{sec:intro}
Transiting planets are crucial for understanding planet formation and evolution, as their masses and radii can be precisely determined. They serve as test beds for transmission spectroscopy studies, providing insights into exoplanet atmospheres (see \cite{Kreidberg2018} and references therein). Initially, Jovian planets were the primary focus of characterization due to their detectability with both radial velocity (RV) and transit methods, \citep{Marcy2005, Fischer2005, Cumming2008, Mayor2011, Char2000} . However, advances in spectrographs on larger telescopes and the advent of space-based transit surveys such as CoRoT \citep{Baglin2006}, \textit{Kepler} \citep{Borucki2010}, and \tess\   \citep{Ricker2015} have led to the predominant discovery of small planets, with radii ranging in size from ones smaller than Neptune's to ones larger than Earth's" ($R_{\rm p} \approx$ 1--4.0\,$R{_\oplus}$).  

Numerous studies have highlighted the abundance of small planets, which now constitute the majority of confirmed exoplanets  \citep{2013ApJS..204...24B, Marcy2005, Howard2010, Howard2012}. \cite{Howard2012} demonstrated 
that the planet occurrence rates per star, for planets that have orbital periods of less than 50 days, decrease from $0.130 \pm 0.008$ to $0.023\pm0.003$ as the planet radius
increases from $2-4~R_{\oplus}$. Similarly, such an occurrence rate decreases to $0.013\pm0.002$ as the planet radius increases to $8-32~R_{\oplus}$. \cite{Mulders2015} observed that small planets are two to three times more common around M dwarfs than FGK-type stars. Another notable trend is the increase in the planet frequency with orbital separation from the host star up to a period of 10 days, beyond which the frequency stabilizes \citep{Howard2012, Catanzarite2011}. 

Studies of small planets have revealed a bimodal radius distribution with a distinct gap, known as the radius valley \citep{Fulton17, Zeng2017, Vaneylen2018, Berger2018}. This gap suggests two distinct populations of exoplanets: one with Earth-like rocky cores with negligible atmospheres and the other with cores and H-He atmospheres. The radius valley, initially observed for solar-type stars, is believed to extend to planets around M dwarfs \citep{VanEylen2020}. Possible explanations for the gap include the photo-evaporation mechanism, whereby X-ray and ultraviolet (XUV) stellar insolation erodes hydrogen and helium atmospheres during formation \citep{Sanz-Forcada2011, Owen2013, Lopez2013}, and the formation of planets in environments with limited gas accretion during their formation process \citep{Owen2013}. Additionally, core-powered mass losses from exoplanets may contribute to the radius valley phenomenon, particularly for young planets in close proximity to their host stars \citep{Ginzburg+2018, Gupta2019, gupta2020}.

In this paper, we present the discovery of TOI-3493~b, a small exoplanet with a mass less than that of Neptune. We describe the properties of the host star in Sect.~\ref{sec:star}, followed by a detailed account of the data utilized for this study, including photometry data, HARPS spectroscopy, and high-resolution imaging data in Sect.~\ref{sec:data}.  Section\,\ref{sec:results}  outlines our methodology and results for determining the planetary and orbital parameters of the system. Finally, in Sect.\,\ref{sec:discussion}, we interpret our findings, and  in Sect.\,\ref{sec:summary} we provide a concise summary.

\section{Host star properties} \label{sec:star}
TOI-3493 is a bright ($V$\,=\,9.32), G1/2V type star \citep{Houk1988}. According to {\cite{GaiaEDR3}}, the star is situated at a distance of about $96.76\pm0.2$~pc. 

For the spectroscopic modeling of TOI-3493, we first analyzed  our high-resolution co-added spectra with the  software SpecMatch-Emp  \citep[SpecM;][]{2017ApJ...836...77Y}, which compares observations to a library of well-characterized stars with effective temperatures between  3000 and  7000~K. The results are listed in Table~\ref{Table: spectroscopic parameters}.
  
\begin{table}
\small
\centering
 \caption{Results from the spectroscopic modeling of TOI-3493 with Spectroscopy Made Easy (SME) 
 and SpecMatch-Emp (SpecM) along with posteriors from the ARIADNE modeling of the SED.}   
\begin{tabular}{llccc }
 \hline
     \noalign{\smallskip}
Method  & $T_\mathrm{eff}$  & $\log g_\star$ & [Fe/H]   &   $V \sin i$    \\  
& (K)  &(dex) &(cgs)       & (km~s$^{-1}$)   \\
    \noalign{\smallskip}
     \hline
\noalign{\smallskip} 
SME  &  $5844\pm 42$  & $4.25 \pm 0.06$ & $0.03 \pm 0.04$   &   $2.2 \pm 0.7$  \\

SpecM   & $5852 \pm 110$  & $4.25\pm 0.12$ & $0.01 \pm 0.09$   &\ldots   \\

ARIADNE  &  $5851\pm 36$   & $4.25\pm 0.07$  & $0.02\pm0.04$   &\ldots   \\ 
  
\hline 
\end{tabular}  \label{Table: spectroscopic parameters}
\end{table}

We then continued with detailed modeling with the
Spectroscopy Made Easy\footnote{\url{http://www.stsci.edu/~valenti/sme.html}.}software \citep[SME;][]{vp96, pv2017}. This code computes synthetic spectra from atomic and molecular line data\footnote{\url{http://vald.astro.uu.se}.} \citep[VALD; ][]{Ryabchikova2015} and stellar atmosphere grids that are fit to the  observations for a chosen set of parameters. We used the SME version 5.2.2 and the Atlas12 atmosphere model \citep{Kurucz2013}. We held the 
micro- and macro-turbulent velocities, $V_{\rm mic}$ and $V_{\rm mac}$, fixed to   1.1~km~s$^{-1}$ \citep{bruntt08} and 3.8~km~s$^{-1}$ \citep{Doyle2014}, respectively. We modeled the spectral lines particularly sensitive to the fit parameters (one at a time). The procedure is described in detail in \citet{2018A&A...618A..33P}.  
The results of the final model are in excellent 
agreement with SpecM and listed in Table~\ref{Table: spectroscopic parameters}. Both SpecM and SME point to a slightly evolved G1 star. 

\begin{figure}[!ht]
 \centering
  \resizebox{\hsize}{!}
            {\includegraphics{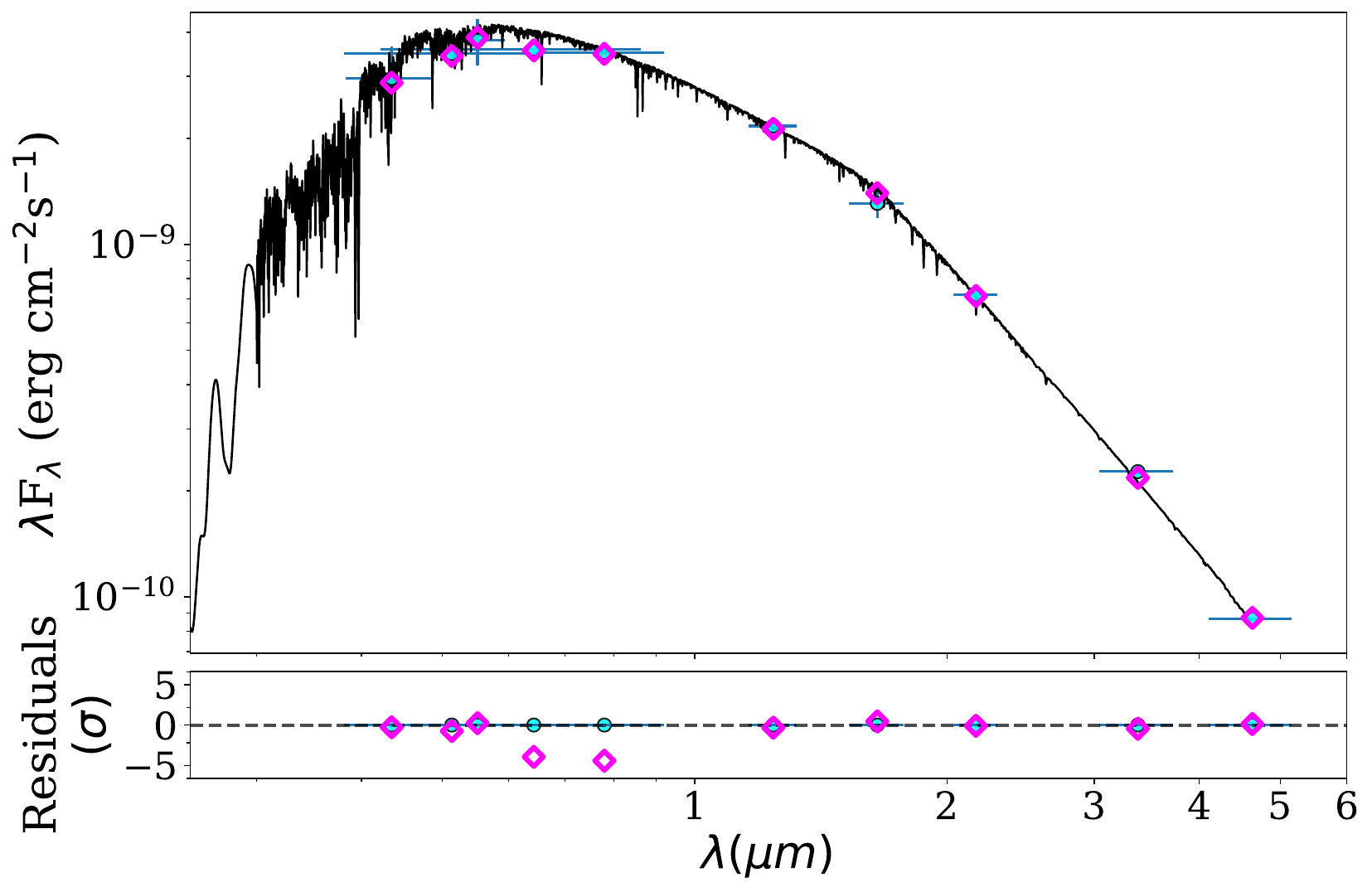}}
   \caption{Spectral energy distribution of TOI-3493 and the best-fitting model 
\citep[{\tt {Phoenix~v2}}, ][]{2013A&A...553A...6H} with the ARIADNE software. The magenta diamonds mark the 
synthetic photometry, while the   observed photometry is shown in blue points.   
The vertical error bars are the 1~$\sigma$ uncertainties of the magnitudes, and the horizontal bars are the effective width of the photometric passbands. The lower panel shows the residuals normalized to the errors of the photometry; hence, the largest relative errors are from \textit{Gaia}~DR3.}
      \label{Figure: SED}
 \end{figure}

To obtain the stellar radius and mass, we used the derived spectroscopic parameters from SME as priors in modeling of the spectral energy distribution (SED) with the 
Python package ARIADNE\footnote{\url{https://github.com/jvines/astroARIADNE}.} \citep{2022MNRAS.513.2719V}. 
We fit the broadband photometry bandpasses Johnson $V$ and $B$ (APASS), $G G_{\rm BP} G_{\rm RP}$ (DR3),    
$JHK_S$ (2MASS), and WISE W1-W2, combined with the \textit{Gaia} DR3 parallax. ARIADNE uses the dust maps of \citet{1998ApJ...500..525S} to put an upper limit of $A_V$. The software also allows fits based on four 
atmospheric model grids. For stars with temperatures above 4\,000~K, ARIADNE uses {\tt {Phoenix~v2}} 
\citep{2013A&A...553A...6H}, {\tt {BtSettl}} 
\citep{2012RSPTA.370.2765A}, 
\citet{Castelli2004}, and 
\citet{1993yCat.6039....0K} atmosphere models. The final radius was computed with Bayesian model
averaging and we found that the {\tt {Phoenix~v2}} has the highest probability (69~\%). 
The final stellar radius becomes $1.23 \pm 0.02~R_\odot$, the luminosity   $1.6~L_\odot$,  and
the extinction is consistent with
zero ($A_\mathrm{V} =0.02\pm0.02$). 
Finally, we obtained the stellar mass and age by  interpolating the MIST \citep{2016ApJ...823..102C} isochrones implemented in ARIADNE and found a stellar mass of $1.02 \pm 0.04~M_\odot$ and an age of $7.3\pm 1.6$~Gyr. Another estimate of the stellar mass was found by combining the radius and $\log g_\star$ which becomes  $0.96\pm 0.15~M_\odot$ in agreement with the mass from the MIST isochrones. We also note that the posteriors for the spectroscopic parameters from the ARIADNE model  (listed in Table~\ref{Table: spectroscopic parameters}) are in excellent agreement with SME and Specmatch-emp. We plotted the SED of TOI-3493 and the resultant best-fitting Phoenix model with the ARIADNE
software, as is shown in Fig.~\ref{Figure: SED}. 

Finally, we used the empirically calibrated gyrochronology relations of \citet{Mamajek:2008} with the rotation period determined from the light curve as discussed in Sec~\ref{subsubsec:Prot}, together with the stellar $T_{\rm eff}$, giving an
estimated system age of $8.5 \pm 0.6$~Gyr, consistent with the age estimated above. Table\,\ref{tab:stellar-param} summarizes the stellar parameters of TOI-3493 with their corresponding uncertainties and references.

\begin{table}
\centering
\tiny
\caption{Stellar parameters of TOI-3493.} \label{tab:stellar-param}
\begin{tabular}{lcr}
\hline\hline
\noalign{\smallskip}
Parameter   & Value             & Reference \\ 
\hline
\noalign{\smallskip}
\multicolumn{3}{c}{\em Name and identifiers}\\
\noalign{\smallskip}
Name    &  HD 119355    & HD1920  \\
TOI     & 3493                  & ExoFOP-TESS \\  
TIC     & 203377303             & Sta18 \\  
\noalign{\smallskip}
\multicolumn{3}{c}{\em Coordinates and basic photometry}\\
\noalign{\smallskip}
$\alpha$ (J2016.0) & 13:43:11.72  & {\em Gaia} DR3  \\
$\delta$ (J2016.0) & -22:35:06.48  & {\em Gaia} DR3 \\
$G$ [mag]        & $9.127\pm0.0027$ & {\em Gaia} DR3 \\
$J$ [mag]       & $8.127\pm0.018$ & Cut03\\
$H$ [mag]        & $7.891\pm0.017$ & Cut03 \\
$K$ [mag]        & $7.776\pm0.026$ & Cut03 \\
\noalign{\smallskip}
\multicolumn{3}{c}{\em Parallax and kinematics}\\
\noalign{\smallskip}
$\pi$ [mas]        & $10.29\pm0.02$ & \textit{Gaia} DR3 \\
$d$ [pc]        & $96.76\pm0.2$ & \textit{Gaia} DR3 \\
$\mu_\alpha \cos \delta$ [mas a$^{-1}$] & $43.32\pm0.02$ & \textit{Gaia} DR3 \\
$\mu_\delta$ [mas year$^{-1}$] & $-72.24\pm0.015$ & \textit{Gaia} DR3 \\
\noalign{\smallskip}
\multicolumn{3}{c}{\em Photospheric parameters and spectral type}\\
\noalign{\smallskip}
Sp. type         & G1/2\,V  & Houk88 \\
$T_{\mathrm{eff}}$ [K]  & $ 5844\pm42$ & This work   \\
$\log g$                & $4.25\pm0.06$ & This work   \\
{[Fe/H]}                & $0.03\pm0.04$ & This work   \\
$v \sin i$ [$\mathrm{km\,s^{-1}}$]    & $2.2\pm0.7$ & This work \\
\noalign{\smallskip}
\multicolumn{3}{c}{\em Stellar properties}\\
\noalign{\smallskip}
$M_\star$ [$M_{\odot}$]       & $1.023\pm0.041$ & This work \\
$R_\star$ [$R_{\odot}$]       & $1.228\pm0.017$ & This work \\
age  & $7.3\pm1.67$ & This work \\
$P_{\rm rot;GP}$ [d] & $37.4\pm1.5$ & This work$^a$ \\
\noalign{\smallskip}
\hline
\end{tabular}
\tablebib{
   \textit{Gaia} DR3: \citet{Gaia2020};
    HD1920: \citet{HDCat1920};
    Houk88: \citet{Houk1988};
    EXOFOP: \url{https://exofop.ipac.caltech.edu/};
    Sta18: \citet{Stassun2018};
    Sta19: \citet{Stassun2019}. 
}
   \tablefoot{
   \tablefoottext{$^a$}{See Sect.~\ref{subsubsec:Prot} for a $P_{\rm rot}$ determination.}
   }
\end{table}

\section{Observations and data acquisition}
\label{sec:data}
\subsection{Transit detection}
\subsubsection{\tess\   photometry} \label{subsubsec:phot-TESS}

\begin{figure*}[!ht]
    \centering
    \includegraphics[width=\textwidth]{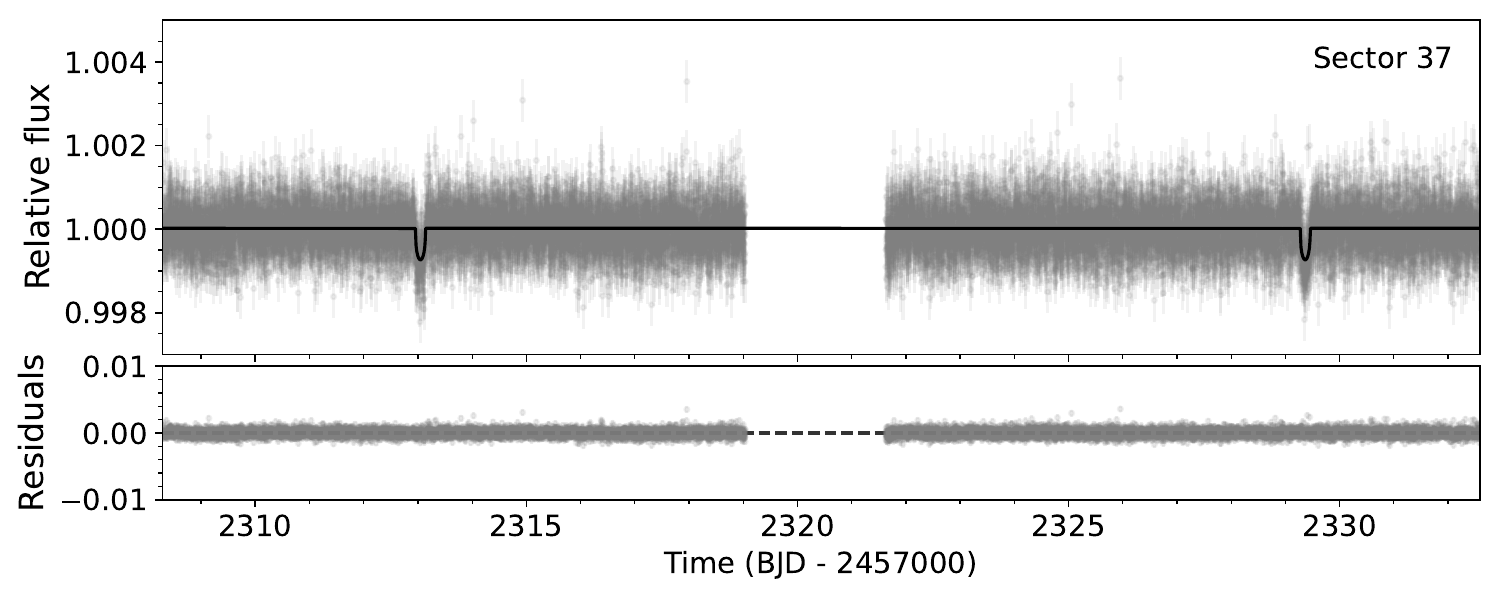}
    \includegraphics[width=\textwidth]{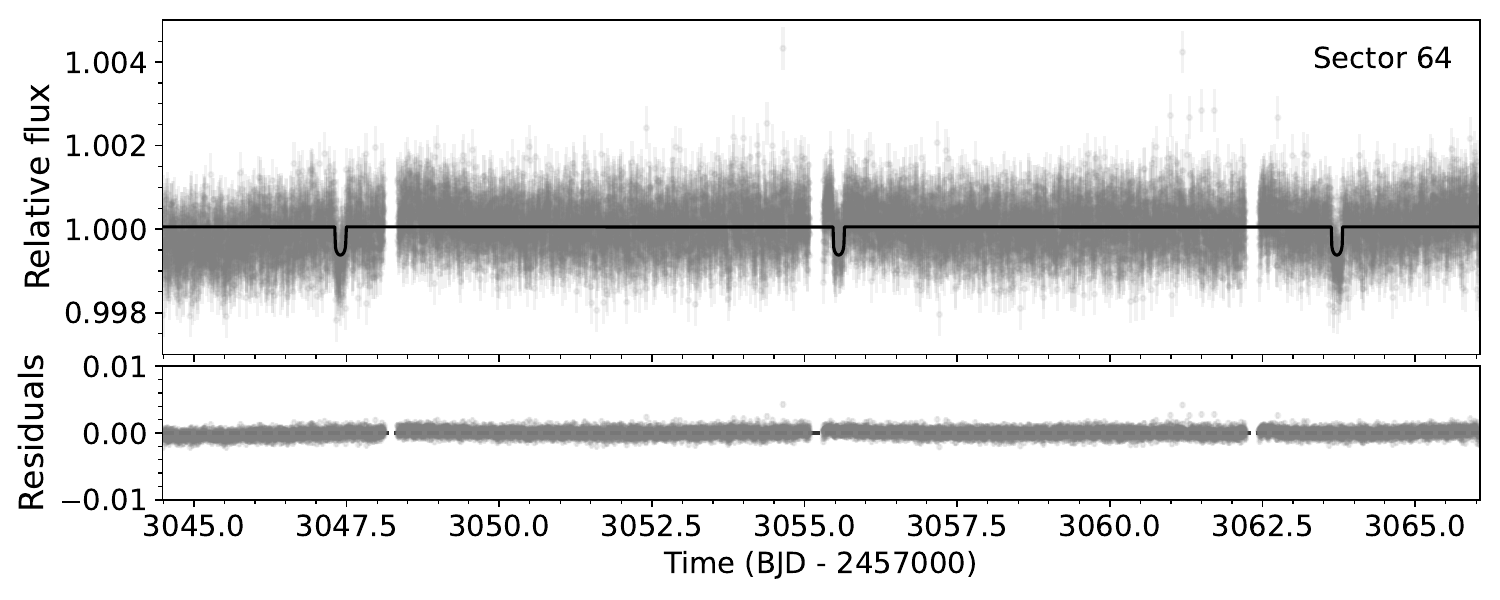}
    \caption{\tess\   PDCSAP light curve for TOI-3493 (gray points) for two sectors, 37 (top) and 64 (bottom), overplotted with the transiting-planet model in black.}
    \label{fig:photo-TESS}
\end{figure*}

\begin{figure}[!ht]
    \centering
    \includegraphics[width=0.5\textwidth]{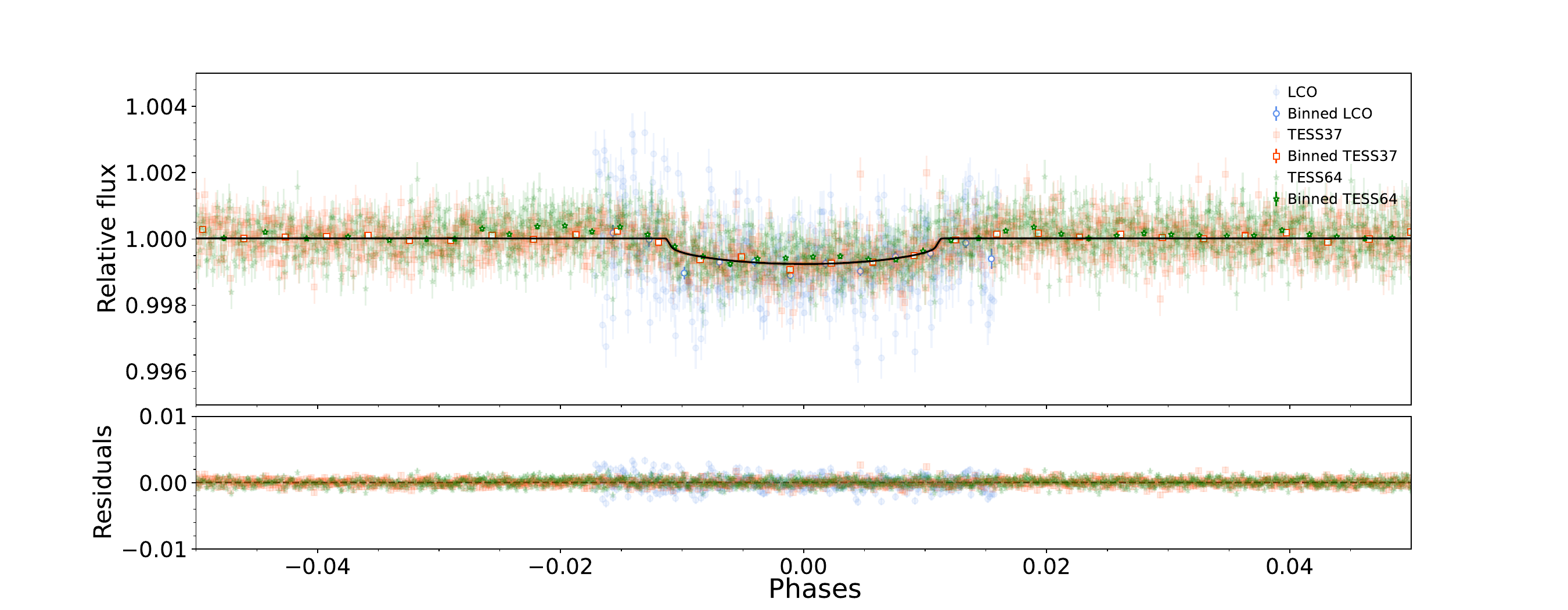}
    \caption{Phase-folded \tess\  and LCOGT transit light curves for TOI-3493~b at 8.15~d.
    Gray circles are LCO observed points, red squares and green stars are 2~minute cadence \tess\   data for sectors 37 and 64, respectively, and darker shades of the same color are binned data for respective datasets (shown only for reference; unbinned points were used to fit the model).
    The best-fit,\texttt{juliet} model (black line; see Sect.~\ref{sec:results}) is overplotted for TOI-3493~b.}
    \label{fig:photo-TESS_phase}
\end{figure}

The Transiting Exoplanet Survey Satellite (\tess) mission, equipped with four 2k~$\times$~2k CCD cameras, each covering a $24~\times~24$~deg$^2$ field of view, has been pivotal in the discovery of numerous exoplanet candidates spanning a range of sizes and orbital separations. The Mikulski Archive for Space Telescopes (MAST)\footnote{\url{https://mast.stsci.edu}, \url{https://archive.stsci.edu/}.} is tasked with archiving both raw and processed light curves pertaining to the observed candidates.

TOI-3493, also designated TIC 203377303, was observed at a 2 minute cadence in Sector 37 in April 2021 and the image data were reduced and analyzed by the Science Processing Operations Center (SPOC; \citealt{SPOC}) at NASA Ames Research Center.  The data validation report \citep{Twicken:DVdiagnostics2018, Li2019} attributed a \tess\   magnitude of 8.7 to the candidate.  According to the difference image centroid tests from the Sector 37 -  Sector 64 multisector search, the host star is located within $1.93 +/-2.60$ arcsec of the source of the transit signal source of TOI-3493. 

The SPOC conducted a transit search of Sector 37 on 4 June, 2021, with an adaptive noise compensating filter \citep{Jenkins2002,Jenkins2010,Jenkins2020}, producing a TCE for which an initial limb darkened transit model was fit \citep{Li2019} and a series of diagnostic tests were conducted to help make or break the planetary nature of the signal \citep{Twicken:DVdiagnostics2018}. The \tess\   Science Office (TSO) reviewed the vetting information and issued an alert on 21 June, 2021 \citep{Guerrero2021}. The \tess\  SPOC (\citealt{SPOC}) identified a transit signal with a periodicity of 16.32~d. Each \tess\  sector observes its designated field of view for approximately 27~d, allowing for a data downlink. However, during sector 37, a data pause lasting 1.19~d occurred, during which a transit of our target took place. This anomaly led to an erroneous determination of the orbital period in sector 37 data, a discrepancy rectified by additional observations conducted during sector 64 in April 2023. The signal was also recovered as additional observations were made in sector 64, and the transit signature passed all the diagnostic tests presented in the data validation reports. Subsequent analysis revised the orbital period to 8.157~d, with a transit depth of 763.88 parts per million (ppm) and a transit duration of 4.4~h, corresponding to a planetary radius of $3.39~R_{\oplus}$ for TOI-3493~b.

We extracted data for SPOC pre-search data conditioning simple aperture photometry \citep[PDCSAP;][]{Smith2012, Stumpe2012, Stumpe2014} for TOI-3493 from sectors 37 and 64. The planet's detection at an 8.17~d period is confirmed in Fig.~\ref{fig:photo-TESS}. Phase-folded data and the best-fit model for the planet in both sectors are presented in Fig.~\ref{fig:photo-TESS_phase} (see Sect.,\ref{subsubsec:joint} for a detailed analysis description).

\begin{figure*}[!ht]
    \centering
    \includegraphics[width=0.4\textwidth]{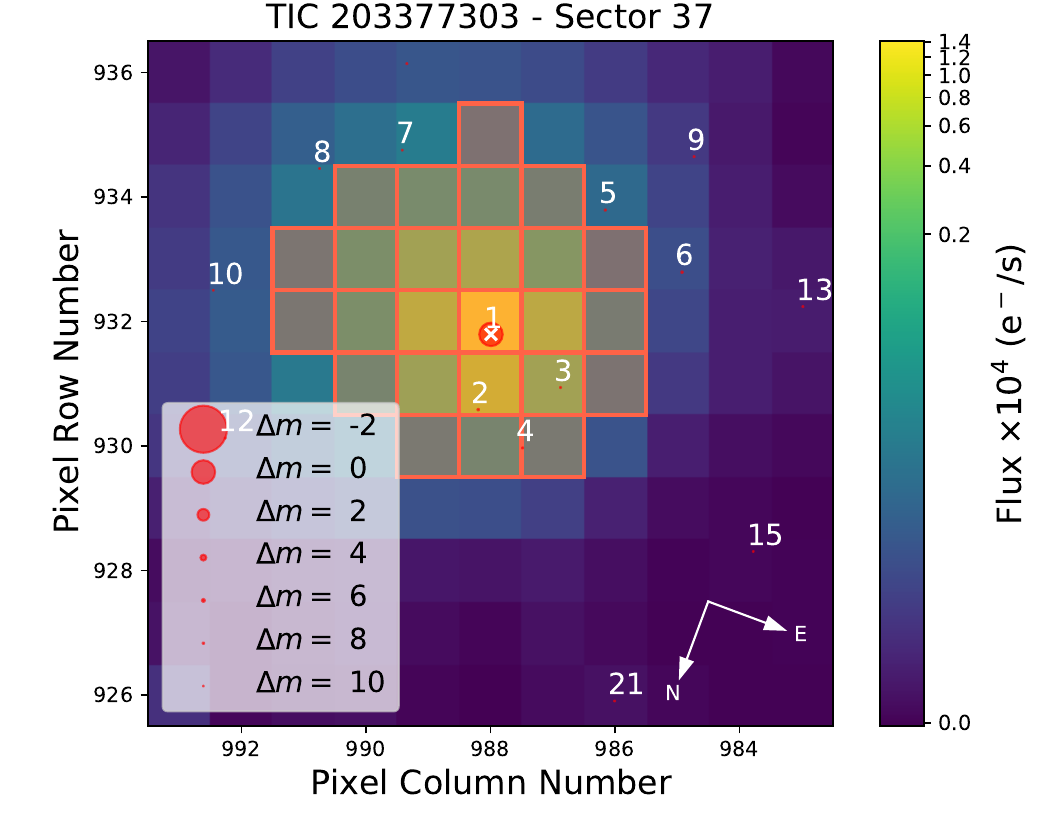}
     \includegraphics[width=0.4\textwidth]{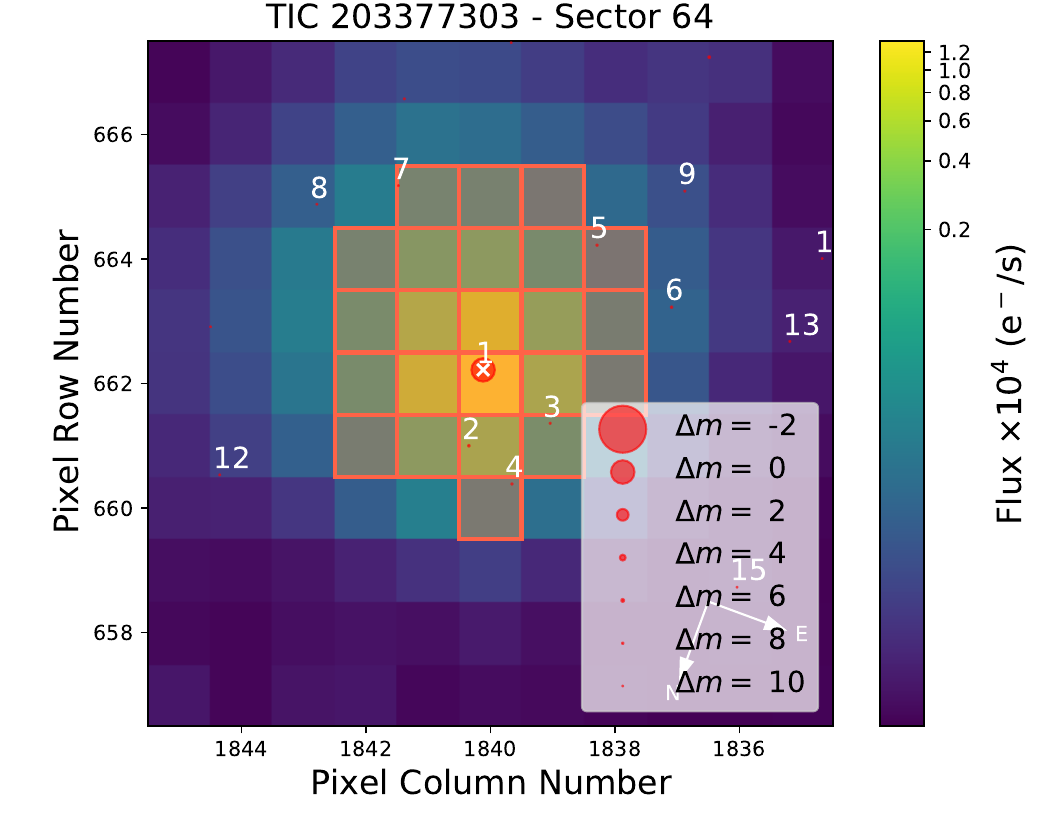}
         \caption{Target pixel files 
         of TOI-3493 in \tess\  sectors 37 (left) and 64 (right). The electron counts are color-coded. The red-bordered pixels are used in the simple aperture photometry. TOI-3493 is represented with a circle marked with $\#$1. The other stars are represented with red circles and the size of the circle related to the magnitude difference. The other stars in the field are relatively faint, and hence appear as tiny dots.}
        \label{fig:TPF}
\end{figure*}

Given the relatively large pixel size of $\sim21~\arcsec$ in the \tess\  imagery, there exists a possibility of contamination from nearby stars affecting the host star. In particular, the {\em Gaia} $G_{RP}$ band ($630-1050$~nm) and the \tess\  $T$ band ($600-1000$~nm) share overlapping wavelength coverage. To assess potential contamination, we overlaid all nearby {\em Gaia} sources against the \tess\   aperture as a reference for the observed field of view, as is shown in Fig.~\ref{fig:TPF}. Notably, no sources exhibit a magnitude difference within 6 of that of TOI-3493, with only a few stars displaying a magnitude contrast of 8 or higher. The TIC contamination of the source is estimated at $\sim0.43\%$, indicating that more than $99\%$ of the flux within the photometric aperture originates from the source. In addition, we performed high-resolution imaging analysis, described in Sect.~\ref{subsec:imaging}, to mitigate the influence of nearby neighbors.

\begin{table}
\centering
\small
\caption{\tess\  observations of TOI-3493.} 
\label{tab:phot_TESS}
\begin{tabular}{cccll}
\hline\hline
\noalign{\smallskip}
Sector   &  Camera   & Cycle   & Start date & End date\\
\noalign{\smallskip}
\hline
\noalign{\smallskip}
37  &  1   &   3 &  3 April, 2021   & 28 April, 2021\\
64  &  1  &   6 &  6 April, 2023    & 1 May, 2023\\
\noalign{\smallskip}         
\hline
\end{tabular}
\end{table}

\subsubsection{LCOGT light curve follow-up}
The pixel scale of the \tess\   is approximately $21\arcsec$ per pixel, with photometric apertures typically extending out to roughly $1\arcmin$, often leading to the blending of multiple stars within the \tess\  aperture. To mitigate the possibility of a nearby eclipsing binary (NEB) blend being the source of the \tess\   detection and to attempt on-target event detection, we conducted observations of TOI-3493 as part of the \tess\   Follow-up Observing Program\footnote{\url{https://tess.mit.edu/followup}.} Sub Group 1 (TFOP; \citealt{collins:2019}). All light curve data are accessible on the EXOFOP-TESS website\footnote{\url{https://exofop.ipac.caltech.edu/tess/target.php?id=203377303}.}.

We observed transit windows of TOI-3493~b on 29 June, 2021, UT and 4 March, 2024, UT in the Pan-STARRS $z$-short band using the Las Cumbres Observatory Global Telescope \citep[LCOGT;][]{Brown2013} 1.0\,m network node at Cerro Tololo Inter-American Observatory in Chile (CTIO). The images were calibrated by the standard LCOGT {\tt BANZAI} pipeline \citep{McCully:2018} and differential photometric data were extracted using {\tt AstroImageJ} \citep{Collins:2017}.

TOI-3493 was intentionally saturated on 29 June, 2021, UT to better measure the light curves of the fainter nearby stars to improve the check for an NEB. We extracted light curves for all eight \textit{Gaia} DR3 and TICv8 neighboring stars within $2\farcm5$ of TOI-3493 that are bright enough in \tess\   band to produce the \tess\   detection (allowing for the magnitude to be an extra 0.5 fainter in the \tess\  band in an attempt to accommodate uncertainties). We thus checked all stars down to 8.3 magnitudes fainter than TOI-3493 (i.e., down to 17.0 mag in \tess\  band). We calculated the root mean square (RMS) of each of the eight nearby star light curves (binned in 5 minute bins) and found that the values are smaller by at least a factor of 5 compared to the expected NEB depth in each respective star. We then visually inspected the light curves to ensure no obvious deep eclipse-like signal existed. We therefore rule out a NEB blend as the cause of the TOI-3493~b detection in the  \tess\  data.

The exposure time for the observation on 4 March, 2024, UT was optimized for TOI-3493 photometry and was mildly defocused in an attempt to improve photometric precision for the purposes of detecting the shallow transit event on TOI-3493. Circular photometric apertures with radius $7\farcs8$ (20 pixels) were used to extract the TOI-3493 differential light curve using eight reference stars. The apertures exclude the flux from all neighbors reported in the \textit{Gaia} DR3 catalog. Telescope guiding was not precise during the observation, so target star centroid location linear detrending was applied, which removed most of the light curve trend. Target star full-width-at-half-maximum (FWHM) and sky-background detrending were also justified by the Bayesian information criterion and were applied to remove further transit model residuals. The transit detection from LCOGT along with the \tess\  data is shown in Fig.~\ref{fig:photo-TESS_phase}.

\begin{table*}
\centering
\tiny
\caption{Ground-based observations of TOI-3493 transits.}
\label{tab:ground-based-transits}
\label{tab:phot}
\begin{tabular}{llllcccccl}
\hline
\hline
\noalign{\smallskip}
Telescope   & Camera or  & Filter & Pixel scale & PSF$^a$ & Aperture & Date & Duration \\
 ~ & instrument & ~ & (arcsec) & (arcsec) & radius (pixel) & (UT) & (min)  \\
\noalign{\smallskip}
\hline
\noalign{\smallskip}
\noalign{\smallskip}

LCO-SSAO (1.0\,m) & Sinistro &  $z_s$ & 0.389 & 4.03 &  18.0 & 05-04-2023 &       363.0  \\

LCO-SAAO (1.0\,m) & Sinistro &  $z_s$ & 0.389   & 3.17  & 17.0 &        04-04-2023 &       537.0  \\

LCO-SSO (1.0\,m) &      Sinistro &      $z_s$ & 0.389 & 3.3 &   16.0 & 19-03-2023 &       463.0 \\

LCO-CTIO (1.0\,m)       &       Sinistro &      $z_s$ & 0.389 & 1.7 &   12.0 & 29-06-2021 &  258.0 \\

\noalign{\smallskip}         
        \hline
    \end{tabular}
   \tablefoot{
   \tablefoottext{$^a$}{Estimated point spread function.}}
\end{table*}

\subsection{High-resolution spectroscopy }\label{subsec:rv}

We used the High Accuracy Radial velocity Planet Searcher (HARPS) spectrograph \citep{Mayor2003} to measure precise RVs for the target. A total of 83 high-resolution spectra, spanning from 16 February, 2022, to 17 August, 2023 (UT), were acquired as part of our extensive observing campaigns 106.21TJ.001 and 1102.C-0923, dedicated to the follow-up of \tess\  transiting planets (PI: D. Gandolfi). These spectra were obtained at a resolving power of R~$\approx$\,115000, covering the wavelength range of 378--691\,nm, with exposure times ranging between 1200 and 1800~s, contingent upon prevailing sky and seeing conditions. The median signal-to-noise ratio (S/N) of the processed spectra is approximately 86 per pixel at 550~nm.

Data reduction of the HARPS spectra was carried out using the dedicated data reduction software DRS \citep{Lovis2007,Pepe2002}, facilitating the extraction of absolute RV measurements by cross-correlating the echelle spectra of TOI-3493 with a G2 numerical mask \citep{Baranne1996}. Furthermore, the DRS was utilized to derive the Ca~{\sc II} H~$\&$~K lines activity indicator (log~R$^\prime_\mathrm{HK}$), as well as the bisector inverse slope (BIS), FWHM, and contrast of the cross-correlation function (CCF). Relative RV measurements and supplementary activity indicators (H$\alpha$ and Na~D lines) were also extracted by employing the Template Enhanced Radial velocity Re-analysis Application (TERRA; \citealt{Anglada2012}).

The absolute DRS and relative TERRA RV measurements of TOI-3493, along with the corresponding activity indicators, are presented in Table~B.1. One HARPS spectrum\footnote{Root file name: HARPS.2023-02-03T07:31:42.961.} exhibited an anomalously low S/N at wavelengths below $\sim$450~nm, likely attributable to misalignment of the HARPS fiber relative to the stellar photo center during exposure. Consequently, this spectrum was deemed unsuitable for analysis, and its associated RV and activity index values are omitted from Table~B.1.

\subsection{High-resolution imaging} \label{subsec:imaging}

The {\em Gaia} DR3 renormalized unit weight error ({\tt RUWE}) value for TOI-3493 is 1.1, much below 1.4, which is considered a critical value to distinguish between single and multiple sources \citep{2018A&A...616A..17A,2018A&A...616A...2L}.
We used the high-resolution imaging facilities at our discretion to confirm the nature of the target source, as is described in the following subsections.

\subsubsection{Gemini} \label{gemini} 

\begin{figure}[!ht]
    \centering
\includegraphics[width=0.45\textwidth]{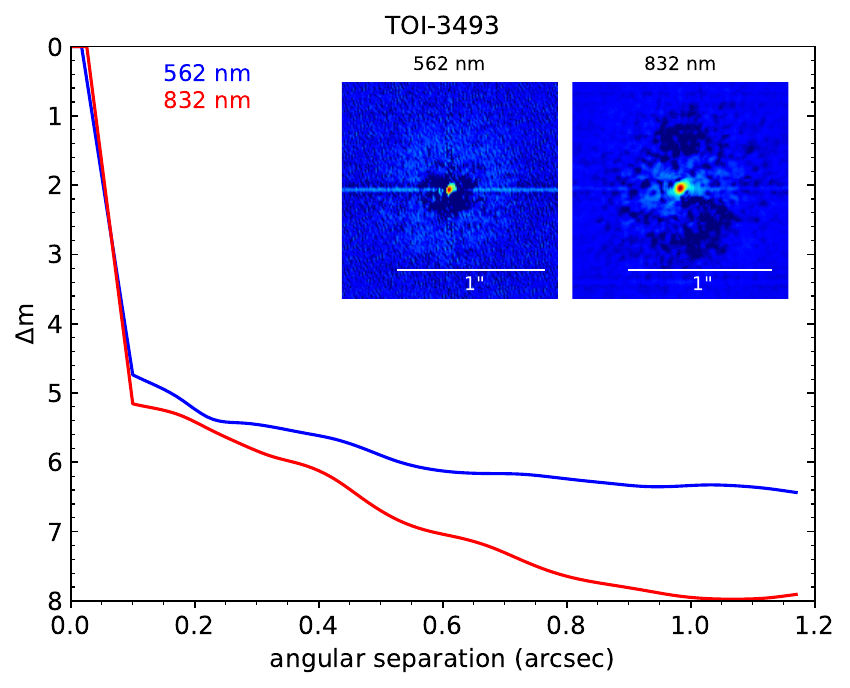}
      \caption{Reconstructed image of TOI-3493 from Zorro speckle instrument on the Gemini South 8-m telescope. The figure includes the 5$\sigma$ contrast curves for the simultaneous observations at 562~nm (blue) and 832~nm (red).}
        \label{fig:speckle}
\end{figure}

If an exoplanet host star has a spatially close companion, that companion (bound or line of sight) can create a false-positive transit signal if it is, for example, an
eclipsing binary (EB) or another form of variable star. ``Third-light” flux from the close companion star leads to an underestimated planetary radius if not accounted for in the transit model \citep{Ciardi2015} and causes non-detections of small planets residing with the same exoplanetary system \citep{Lester2021}. Additionally, the discovery of close, bound companion stars, which exist in nearly one half of FGK-type stars \citep{Matson2018} and less frequently for M-type stars, provides crucial information for our understanding of exoplanetary formation, dynamics, and evolution \citep{Howell2021}. Thus, to search for close-in bound companions that are unresolved in \tess, we obtained high-resolution speckle imaging observations of TOI-3493. TOI-3493 was observed on May 17, 2022, UT using the Zorro speckle instrument on the
Gemini South 8-m telescope\footnote
{\url{https://www.gemini.edu/sciops/instruments/alopeke-zorro/}.}.  Zorro provides simultaneous speckle imaging in two bands (562nm and 832 nm) with output data products including a reconstructed image with robust contrast limits on companion
detections \citep{Scott2021, Howell2022}. Three sets of $1000\times 0.06$~s exposures were collected and subjected to Fourier analysis in our standard
reduction pipeline (see \citealt{howell2011}). Figure~\ref{fig:speckle} shows our final contrast curves and the reconstructed speckle images. We find that TOI-3493 is a
single star revealing no close companions brighter than 5-8 magnitudes below that of the target star from the diffraction limit (20 mas) out to $1.2~\arcsec$. At the distance of TOI-3493 (d=97 pc), these angular limits correspond to spatial limits of 2 to 116~au.

\subsubsection{SOAR}
\begin{figure}[!ht]
    \centering
\includegraphics[width=0.45\textwidth]{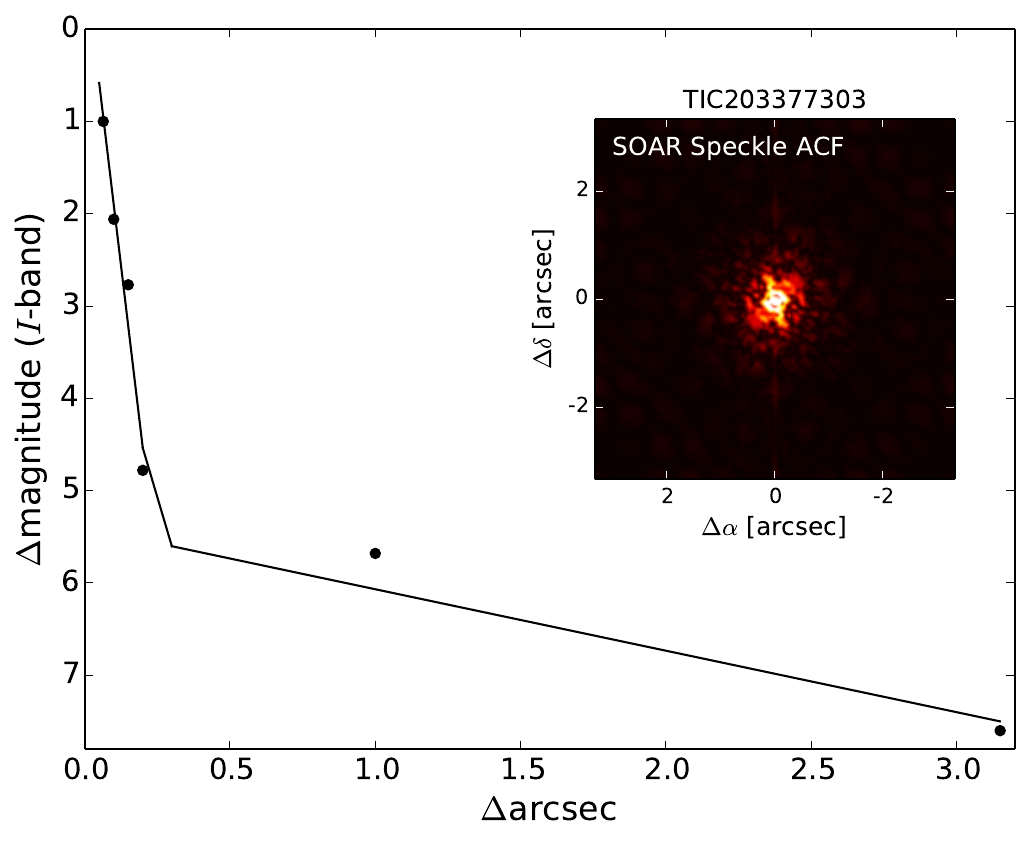}
      \caption{SOAR image of TOI-3493 and contrast curve ($5 \sigma$ limits in black dots, rms dispersion in magenta)}
        \label{fig:soar}
\end{figure}

High-angular-resolution imaging is needed to search for nearby sources that can contaminate the \tess\  photometry, resulting in an underestimated planetary radius, or
that can be the source of astrophysical false positives, such as background eclipsing binaries. We searched for stellar companions to TOI-3493 with speckle imaging on the
4.1-m Southern Astrophysical Research (SOAR) telescope \citep{Tokovinin2018} on 14 July, 2021, UT, observing in the Cousins I band, a similar visible bandpass to \tess. This
observation was sensitive to a 5.7-magnitude fainter star at an angular distance of $1~\arcsec$ from the target. More details of the observations within the SOAR \tess\ 
survey are available in \cite{Ziegler2020}.
The $5\sigma$
detection sensitivity and speckle autocorrelation functions from the observations
are shown in Fig.~\ref{fig:soar}. No nearby stars were detected within $3\arcsec$ of TOI-3493 in
the SOAR observations.

\section{Analysis and results} \label{sec:results}

\subsection{Investigating the periodic signals}
\label{subsubsec:periodogram}

The periodic signals evident in both the RV and photometry data of the star may arise from intrinsic stellar activity, such as dark spots, bright faculae, and magnetic fields, which manifest as periodically modulated signals in long baseline photometry data. Consequently, we conducted separate investigations of the spectroscopic and photometric data.

\subsubsection{Photometry data}
\label{subsubsec:Prot}
To probe stellar activity induced phenomena, we analyzed archival photometry data obtained from the SuperWASP mission, accessible through the SuperWASP consortium via the NASA Exoplanet Archive\footnote{\url{https://exoplanetarchive.ipac.caltech.edu/docs/SuperWASPMission.html}.}. The SuperWASP project comprises two robotic observatories situated in the northern and southern hemispheres, each equipped with eight cameras boasting a broad field of view of $7.8 \times 7.8$~deg$^2$ per camera \citep{Butters2010}. TOI-3493 was included in the photometric observations conducted by SuperWASP-South, situated at the South African Astronomical Observatory (SAAO). These observatories are designed to conduct systematic sky monitoring for photometric phenomena. Observations of TOI-3493 spanned from May 2006 to August 2014, yielding approximately 10,000 measurements. The data were acquired using a broadband filter with a passband spanning from 400~nm to 700~nm.

\begin{figure*}[!ht]
    \centering
\includegraphics[width=0.33\textwidth]{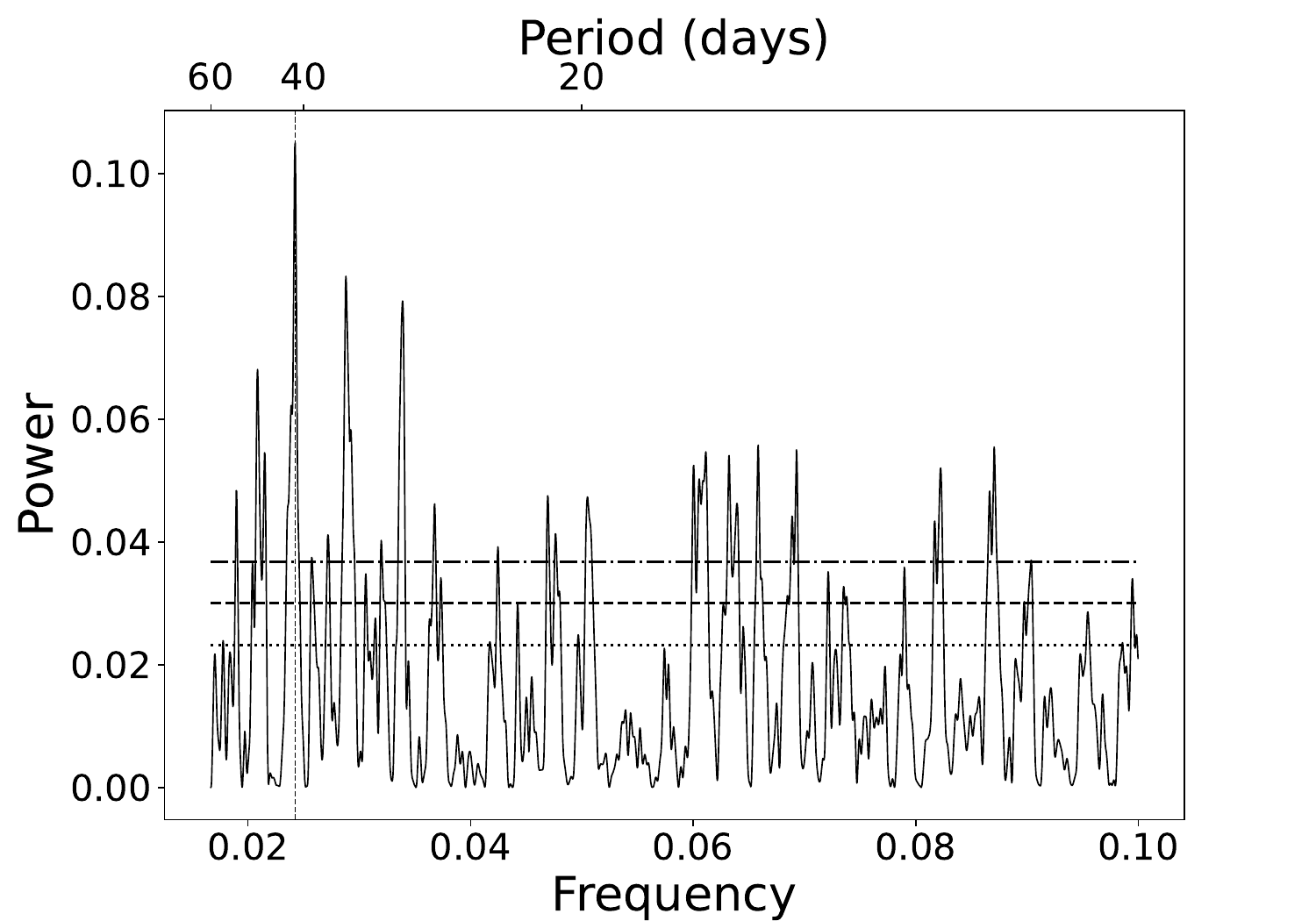}
    \includegraphics[width=0.33\textwidth]{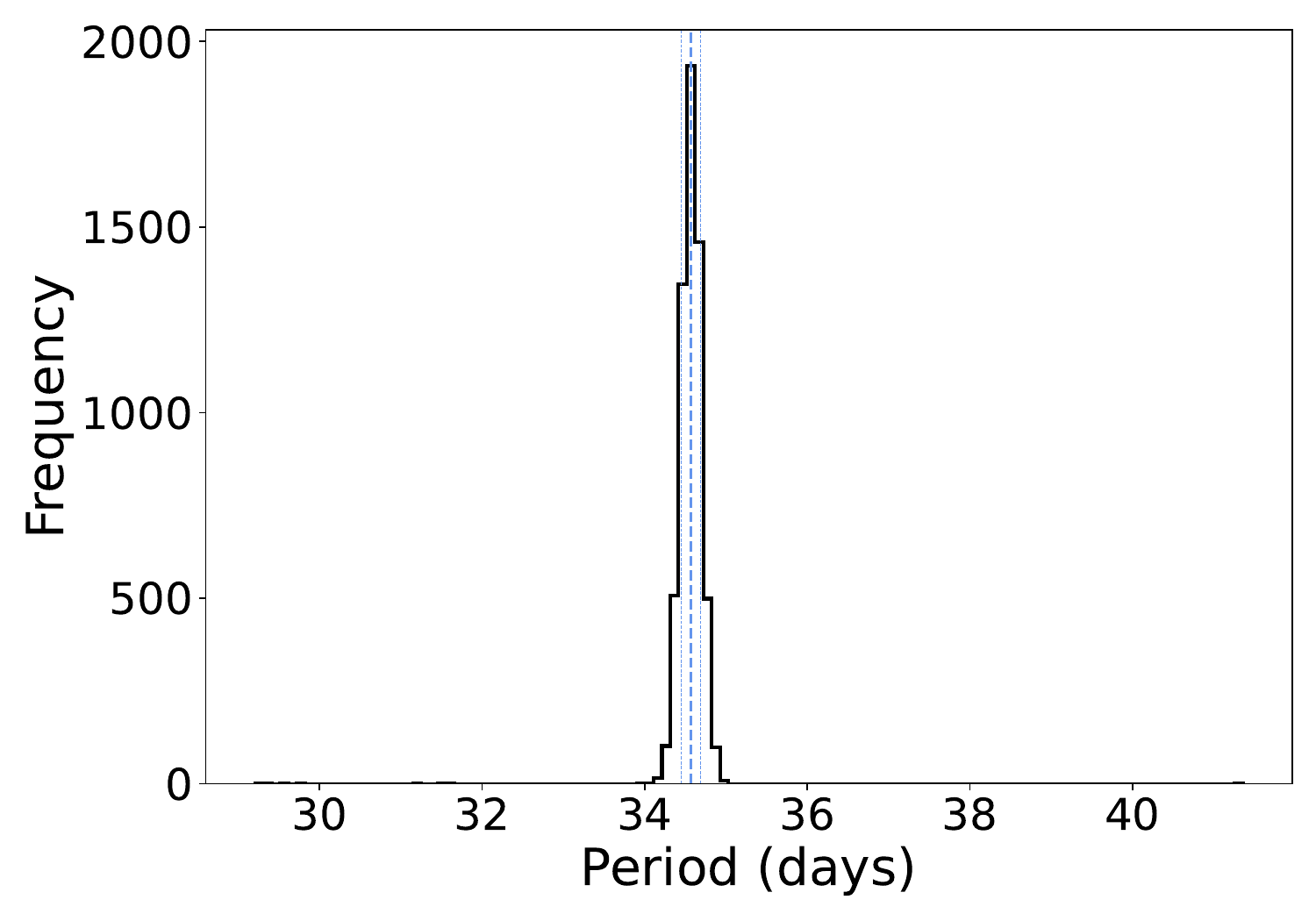}
    \includegraphics[width=0.33\textwidth]{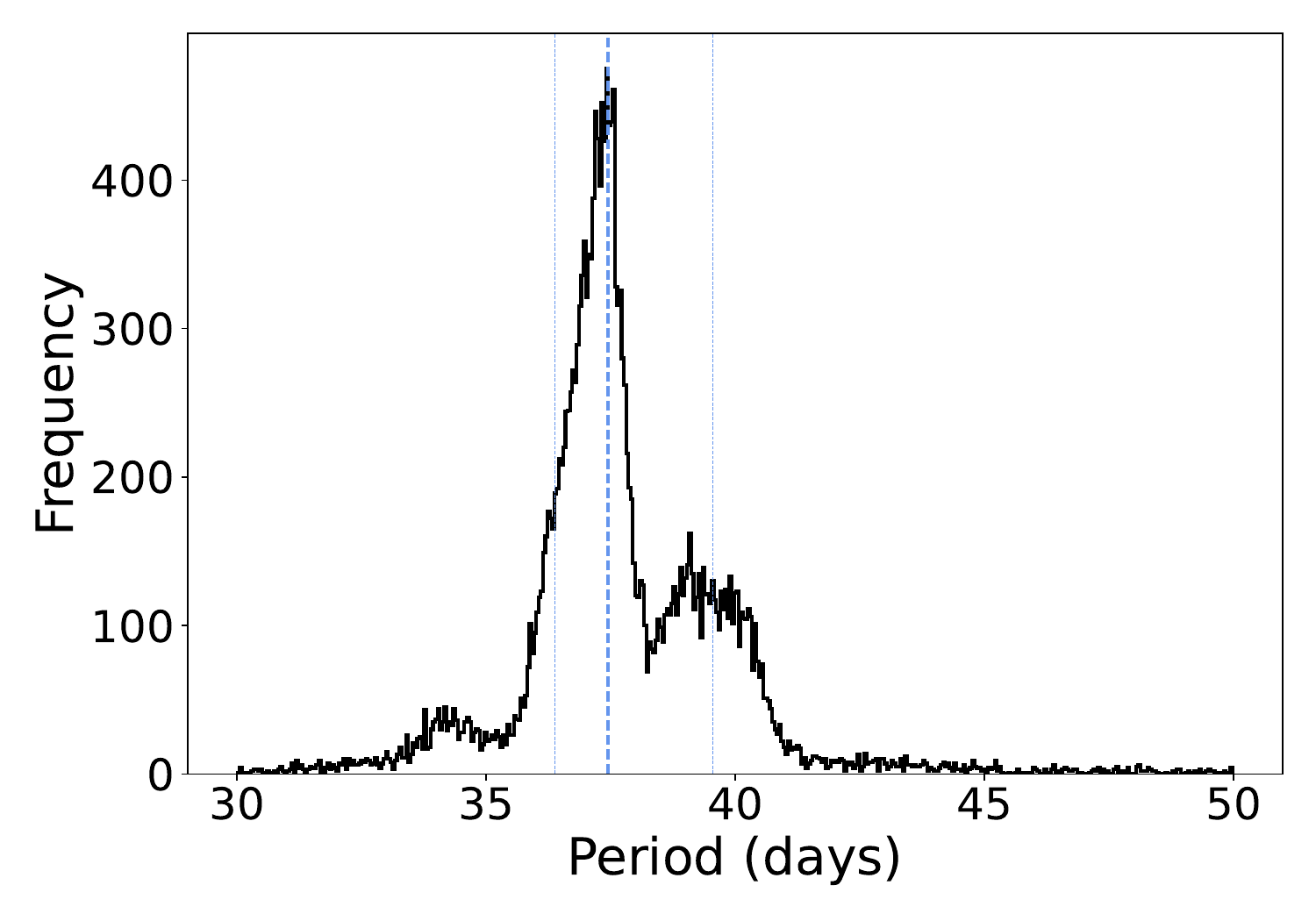} 
    \caption{Rotational period of TOI-3493 determined through photometry and spectroscopy data described in the following panels. (Left) GLS periodogram for the SuperWasp data plotted, with a significant peak occurring at 41~d and several secondary peaks occurring between 30--50~d. The false alarm probabilities at $0.1\%$, $1\%$, and $10\%$ are plotted from top to bottom. (Middle) PDF for the most significant period at 34.5~d by modeling the nightly binned photometric SuperWASP data with GPs. (Right) PDF for the estimated rotational period of the star, while modeling the RV data with GPs. The PDF has a peak at $37.4\pm1.5$~d followed by a small peak centered at 39~d.}
    \label{fig:GP-Prot}
\end{figure*}

We conducted a generalized Lomb-Scargle (GLS) periodogram analysis \citep{Zechmeister2009}
to the SuperWasp data. The GLS periodogram shows a strong peak at 41~d, with subsequent peaks occurring at 30, 34, and 47~d, as is shown in the left panel of Fig.~\ref{fig:GP-Prot}. Due to a large number of significant peaks occurring in the periodogram, we employed a Gaussian process (GP) model, implemented via the fitting tool \texttt{juliet} \citep{juliet}, to analyze the SuperWASP light curves. The data were nightly binned to mitigate short-period variations, and we utilized the built-in quasi-periodic (QP) kernel of the Python library \texttt{george} \citep{Ambikasaran2015}. Mathematically, this kernel combines an exponential-squared component with a sinusoidal component, expressed as
\begin{equation}\label{eq:QP-GP}
k(\tau)=\sigma^{2}{\rm GP}\exp\left(-\alpha{\rm GP}\tau^2 -\Gamma \sin^2\left[\frac{\pi \tau}{P_{\rm rot;GP}}\right] \right)
,\end{equation}
where $\sigma_{\rm GP}$ denotes the GP amplitude (in ppm), $\Gamma$ represents the dimensionless amplitude of the GP sine-squared component, $\alpha$ indicates the inverse length scale of the GP exponential component (in days$^{-2}$), $\tau$ signifies the time lag (in days), and $P_{\rm rot;GP}$ denotes the rotational period of the star (in days). While the sinusoidal signal associated with active regions on the star may vary over time \citep{Angus2018}, the QP kernel offers an effective means of modeling such intricate periodic behaviors.

We treated data from each camera as originating from distinct instruments to avoid potential instrumental offsets across datasets. To account for relative flux offsets between different instruments, we assumed a normal distribution ranging from 0 to $1000$. The uninformative priors were applied to all parameters: $\sigma_{\rm{GP}}$ (uniform distribution between $10^{-8}$~ppm and $10^8$~ppm), $\Gamma$ (uniform distribution between $10^{-6}$ and $10^{6}$), instrumental jitter (uniform distribution between 0.1~ppm and $10^{9}$~ppm), $\alpha$ (uniform between $10^{-10}$~d$^{-2}$ and $1$~d$^{-2}$), and $P_{\rm rot;GP}$ (uniform between 10 and 60~d). The rotational period, determined from the GP analysis, was found to be $P_{\rm rot;GP}$~=~$34.6\pm0.1$~d, as is seen from the peak of the posterior distribution function (PDF) seen in the middle panel of Fig.~\ref{fig:GP-Prot}.

\subsubsection{Spectroscopy data} \label{subsec:RV_GLS}

\begin{figure}[!ht]
    \centering
    \includegraphics[width=0.5\textwidth]{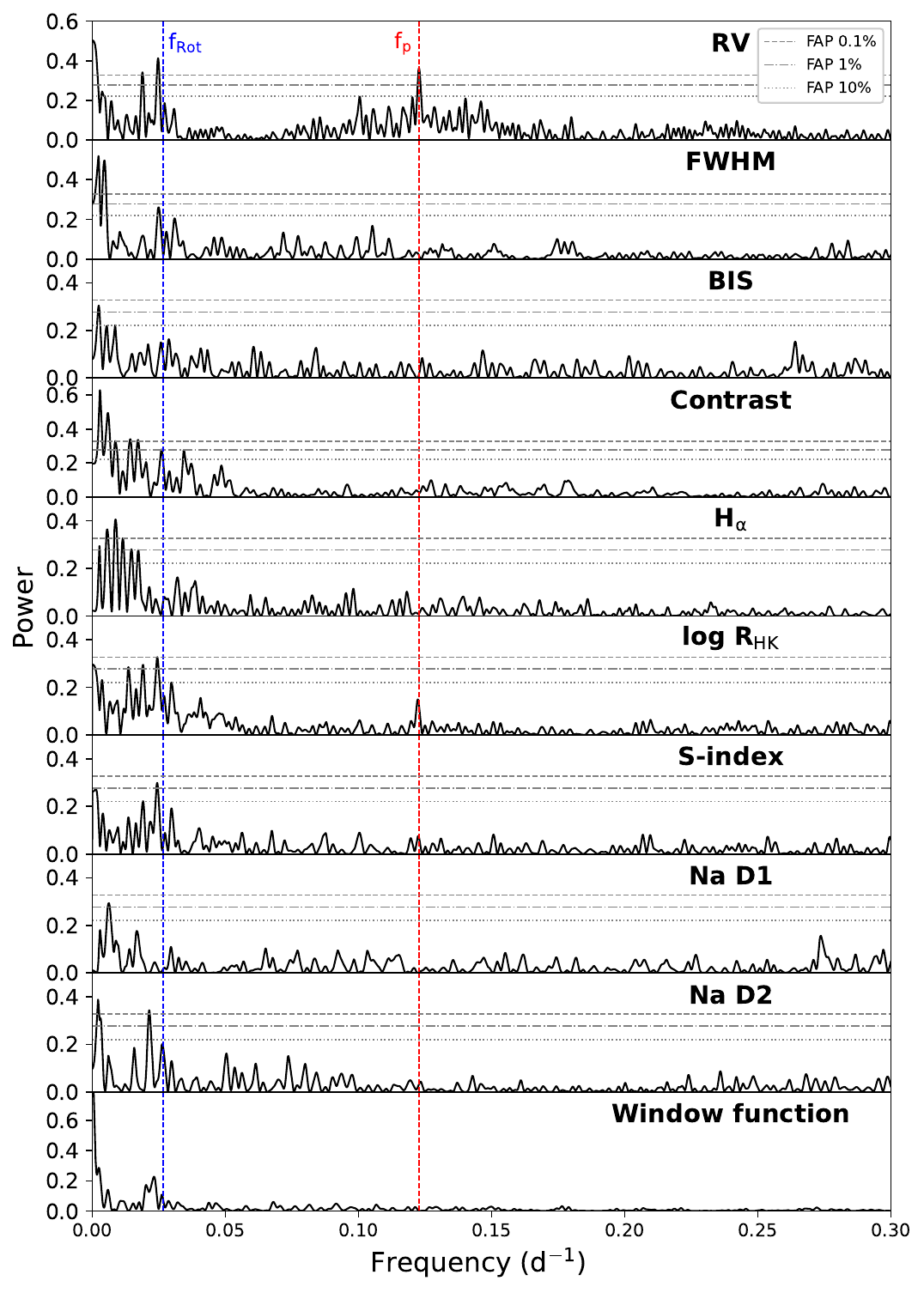}
    \caption{GLS periodograms of various parameters derived from the HARPS data: RV measurements, FWHM of the CCF, BIS, contrast, H$\alpha$ index, log$R'_{HK}$ index, S-index, Na~{\sc i}D$_1$, and Na{\sc i}D$_2$, juxtaposed with the respective window function to scrutinize potential alias signals. A vertical red line is annotated to denote the frequency corresponding to the period of the planet (8.15~d). and the blue line corresponds to the rotational period of the star (37.4~d)}    
    \label{fig:spectral_indices}
\end{figure}

We conducted a GLS periodogram analysis \citep{Zechmeister2009} on the HARPS RV data for TOI-3493, alongside its associated activity components extracted from the data. The GLS periodogram of the RV data along with the activity indicators are displayed in Fig.~\ref{fig:spectral_indices}. Followed vertically by the RV panel on top are the activity indicators encompassing the FWHM of the CCF, BIS, contrast, H$\alpha$ activity indicator, log$R'_{HK}$, S-index, Na~{\sc i}D$_1$, Na{\sc i}~D$_2$, and the window function in each panel, respectively. 

We pre-whitened the data by sinusoidal fitting and subtracting the prominent 8.15 d signal evident in the RV data. Subsequent examination of the RV periodograms, in comparison to other activity indicators, reveals no conspicuous signal at the 8.15~d, accompanied by additional features between 34 and 41~d and in some cases around 50~d. The contrast activity indicator exhibits a prominent signal at half of this period, namely 20~d. Similarly, analysis of other indicators such as H$\alpha$ reveals a peak occurring around 37~d. The activity indices, alongside their associated uncertainties, are detailed in Table~B.1.

While modeling the RV data, as is discussed in Sect.~\ref{appsubsec:rv-juliet}, for its planetary signal, an additional component in the form of GPs was introduced to account for the residual peaks observed in the RV data. These peaks overlap with some of the activity indicator indices shown in Fig.~\ref{fig:spectral_indices}. The period derived from the GP reflects the rotational modulation of the star. The PDF exhibits a primary peak at $37.4\pm1.5$~d, along with a secondary, broader peak around 39~d, as is illustrated in the right panel of Fig.~\ref{fig:GP-Prot}.

The 41~d signal, observed as the strongest peak in the GLS periodogram (left panel), and the 34~d signal, identified through GP modeling of the photometry data (middle panel), are found to be one-year (365~d) aliases of the 37.4~d signal. Therefore, based on a comprehensive analysis of the periodic signals seen in both the photometric and spectroscopic data, we conclude that the rotational period of the star is 
$37.4\pm1.5$~d.

\subsection{Orbital parameters for TOI-3493 system}

Modeling the RV and transit data of a system allows for the determination of crucial parameters, such as the radius, mass, and bulk density of the transiting planet, along with its orbital parameters, such as the inclination and eccentricity. We utilized the Python-based library \texttt{juliet} to model all available transit and RV data. \texttt{juliet} employs the nested sampling algorithm from \texttt{dynesty} \citep{Speagle2020} to compute the Bayesian model log evidence, $\ln \mathcal{Z}$, along with its posterior distribution. This library relies on other Python packages such as \texttt{radvel} \citep{Fulton2018} for RV modeling and \texttt{batman} \citep{Kreidberg2015} for transit modeling. To account for nonperiodic variations, GPs are incorporated into the modeling through packages such as \texttt{george} \citep{Ambikasaran2015} and \texttt{celerite} \citep{Foreman-Mackey2017}. The modeling of the TOI-3493 system was conducted in three steps, outlined below.

\subsubsection{Modeling the transit data only}
\label{phot-juliet}

The \tess\  data for TOI-3493 was available for two \tess\  sectors, 37 and 64, from which the PDCSAP light curves were retrieved. Using the SPOC reported period of 16.32 d and ephemeris as priors, we initially attempted to fit a 16.32~d period to the dataset. However, the RV data indicated a strong peak at 8.15~d instead of 16.32~d, prompting us to refine our transit search at half the quoted period, at 8.15~d. There are systematic gaps in the \tess\  data due to the data downlink time. A transit occurred during one the gaps for the TOI-3493 system resulting in an incorrect reported period. We let the transiting planet have a uniform period prior between 7.5 to 9~d. We chose the same transit center time as mentioned by SPOC but also kept the prior information with a uniform distribution ranging from BJD 2459310.0 to BJD 2459314.0.

The stellar density ($\rho_\star$) was utilized as a free parameter instead of the scaled stellar radius ($a/R_\star$). As was discussed in \cite{Sozzetti2007}, fitting for the $\rho_\star$ is considered a better choice as it incorporates information from both stellar mass and radius. We also allowed the $\rho_\star$ prior to have a uniform distribution with a wide range of possible choices between 100 and 10000 kg\,m\,$^{-3}$. The planet-to-star ratio ($p_{\rm 1}$) and impact parameter ($b_{\rm 1}$) were parameterized based on \cite{Espinoza2018}, with other parameters set within appropriate ranges. The successful execution of the script allowed us to derive the orbital period ($P_{\rm b}$) and central transit times ($t_{0b}$), which were subsequently used as priors for simultaneous data fitting. The parameters that are better determined through RV such as eccentricity-omega (e\,--\,$\omega$) were fixed at 0 and 90$^\circ$, respectively, for this step. In the absence of any close known companions, discussed in Sec~\ref{subsec:imaging}, we fixed the dilution factor to 1.0. We chose quadratic limb darkening law with parameters varying uniformly between 0 and 1. The flux offset had a normal distribution around 0 with a 0.1 width in units of normalized flux. We set the priors for the flux scatter to vary uniformly between $10^{-5}$ to $10^{5}$\,ppm. The successful execution of the script let us derive the orbital period, $P_{\rm b}$, as $8.1594647^{+0.0000584}_{-0.0000554}$~d and the central transit times, $t_{0b}$, as BJD $2459313.0472^{+0.0030 }_{-0.0023}$. These parameters were used as priors when we fit the data simultaneously as is discussed in Sec~\ref{subsubsec:joint}.

\subsubsection{Modeling the RV data only} \label{appsubsec:rv-juliet}

The preliminary analysis of the HARPS RV data, using the GLS periodogram, facilitated the determination of the correct orbital period for the transiting planet orbiting TOI-3493. Contrary to the 16.32~d period reported by the \tess\  mission's SPOC, we identified a period of 8.15~d (\citealt{SPOC}) for the planetary transit. This was corroborated by independent transit modeling efforts detailed in Sect.~\ref{phot-juliet}. Subsequent to the removal of the planetary signal at 8.15~d, a prominent RV signal emerged at 37~d, accompanied by residuals at periods of 34, 41, and 52~d. These residual periods are related to the stellar rotational period and its alias, as is discussed in Sect. \ref{subsubsec:Prot}.

We initiated our modeling process by fitting the transit data assuming a circular orbit, denoted as the “1cp model.” Using the derived values of $P_{\rm b}$ and $t_{0b}$ from the transit-only best-fit (Sect. \ref{phot-juliet}), we varied the semi-amplitude for the RV between 0 and 50~m\,s$^{-1}$ with a uniform distribution. The instrument offset parameter was kept uniform between -10 and 10~m\,s$^{-1}$, and the RV jitter was varied between 0.01 and 10~m\,s$^{-1}$ with a log uniform distribution. For circular (cp) models, the eccentricity ($e$) and argument of periastron ($\omega$) were fixed at 0 and 90, respectively.
In cases in which circular models were not fit, denoted as Keplerian (ep) models, the parameters were parameterized accordingly \citep{juliet}.

We assessed various models against the RV data, employing the selection criteria by \cite{Trotta2008} to discern the optimal model. The difference in Bayesian evidence ($|\Delta \ln \mathcal{Z}|$) between any two models guided the decision-making process, with a threshold of $|\Delta \ln \mathcal{Z}| > 5$ signifying a significant preference, $|\Delta \ln\mathcal{Z}| > 2.5$ indicating moderate favorability, and values $\leq 2.5$ suggesting models as being “indistinguishable,” thus favoring simpler models. Given the presence of significant peaks in the RV data at 8.15~d and around 41~d, we designated the 2cp model as the baseline for comparison. Details of the different models and their Bayesian evidence are presented in Table \ref{tab:bayesian}.

While modeling the RV peaks with circular (cp) or Keplerian (ep) orbits, we consistently observed a preference for cp models over ep models. When considering long-term trends in the data, the 1cp model demonstrated moderate favorability over the 2cp model, with a linear trend significantly outperforming a quadratic trend for the 1cp model. The linear trend in the data indicates radial acceleration of the system with a value of $-0.017\pm{0.002}$ $\mathrm{m\,s^{-1}day^{-1}}$. If this acceleration is due to a third body in the system, as is outlined in \cite{Smith2017} and related references, the inferred planet would have a mass exceeding $3~M_J$ and an orbital separation within 4 AU. However, the current observational time baseline is insufficient to confirm its existence.

The subtraction of the first primary signal seen in the GLS periodogram leaves us with the presence of multiple peaks in the RV data. These periods have an overlapping range with the periods observed for the activity indicators within the range of 34 and 50~d. As in Sec~\ref{subsubsec:Prot}, where GPs are incorporated to compute the photometrically derived rotational period of the star, we believe that similar quasi-periodic variations are causing residual peaks in the RV data. We thereby employed the QP kernel, similarly, to capture complex periodic signals induced by stellar spots on the surface (\citealt{2011A&A...525A.140D, 2014MNRAS.443.2517H}). These signals show their presence not only in the light curve data but also affect certain stellar lines more than others. This results in line shape variations and similar effects showing peaks in the RV data that can be misinterpreted as a planetary signal. We used GP as a tool to model these complex periodic signals. For the GP parameters, we used uniform priors for the GP amplitude ($\sigma_{\rm{GP,RV}}$) between 0 and 100 m\,s$^{-1}$. The inverse length scale of the external parameter ($\alpha_{\rm{GP,RV}}$) and the amplitude of the sine part of the kernel ($\gamma_{\rm{GP,RV}}$) had a log uniform distribution between $10^{-10}$ to 1 and between 0.01 to 10, respectively. We kept the rotational period of the GP ($P_{\rm{rot;GP,RV}}$) varying uniformly between 30 to 60~d to account for the additional peaks seen in the RV data that overlap with some signals seen in the activity indicators. The analyses resulted in the 1cp$+$GP model exhibiting the lowest $\ln{\mathcal{Z}}$ and a significant distinction from our fiducial 2cp model, establishing it as the optimal solution.

\begin{table}
\centering
\small
\caption{Comparison of RV-only models.}
\label{tab:bayesian}
\begin{tabular}{l c c}
\hline\hline
\noalign{\smallskip}
Models & $\ln{\mathcal{Z}}$  & $|\Delta\ln{\mathcal{Z}}|$\\
\noalign{\smallskip}
\hline
\noalign{\smallskip}
\multicolumn{3}{c}{\it Two-signal models (without activity modeling)}\\[0.1cm]
\noalign{\smallskip}
1cp   & $-227.77$ & 7.76 \\
1ep    & $-238.57$ & $18.56$ \\
2cp   & $-220.02$ & $0$ \\
1cp+1ep & $-230.38$ & $10.37$ \\
2ep   & $-233.95$ & $13.94$ \\
1cp+lt        & $-216.75$ & $-3.26$ \\
1cp+qt        & $-229.98$ & $9.96$ \\
2cp+lt        & $-219.09$ & $-0.93$ \\
2cp+qt        & $-219.29$ & $-0.72$ \\
1cp+GP        & $-205.21$ & $-14.80$ \\
1ep+GP   & $-209.53$ &  $-10.49$   \\
\hline
\end{tabular}

\tablefoot{Here, ``cp,'' ``ep,'' lt, qt, and ``GP'' refer to a circular planet model, eccentric planet model, linear trend, quadratic trend, and GPs, respectively. 
$\ln \mathcal{Z}$ and $|\Delta\ln \mathcal{Z}|$ are the log-evidence and relative absolute log-evidence with respect to the fiducial model (2cp here), respectively.
}
\end{table}

\subsubsection{Joint modeling of the RV and transit data}
\label{subsubsec:joint}
Simultaneous modeling of both RV and transit data facilitates a more precise constraint on the planet and orbital parameters. For instance, parameters such as the orbital period ($P$) and transit center time ($T_{c}$) are influenced by both datasets and are best determined when modeled jointly. In our analysis of the TOI-3493 system, we employed both HARPS RV data and \tess\  light curves for joint modeling.

The priors used in this joint modeling step were the same as the ones discussed in the previous individual modeling steps, in Sects. \ref{phot-juliet} and \ref{appsubsec:rv-juliet}. These priors are detailed in Table \ref{tab:priors}. Notably, the RV semi-amplitude ($K$) for the planet was determined to be $2.81^{+0.36}_{-0.37}$\,m\,$\rm s^{-1}$. 

\begin{figure*}[!ht]
    \centering
    \includegraphics[width=\textwidth]{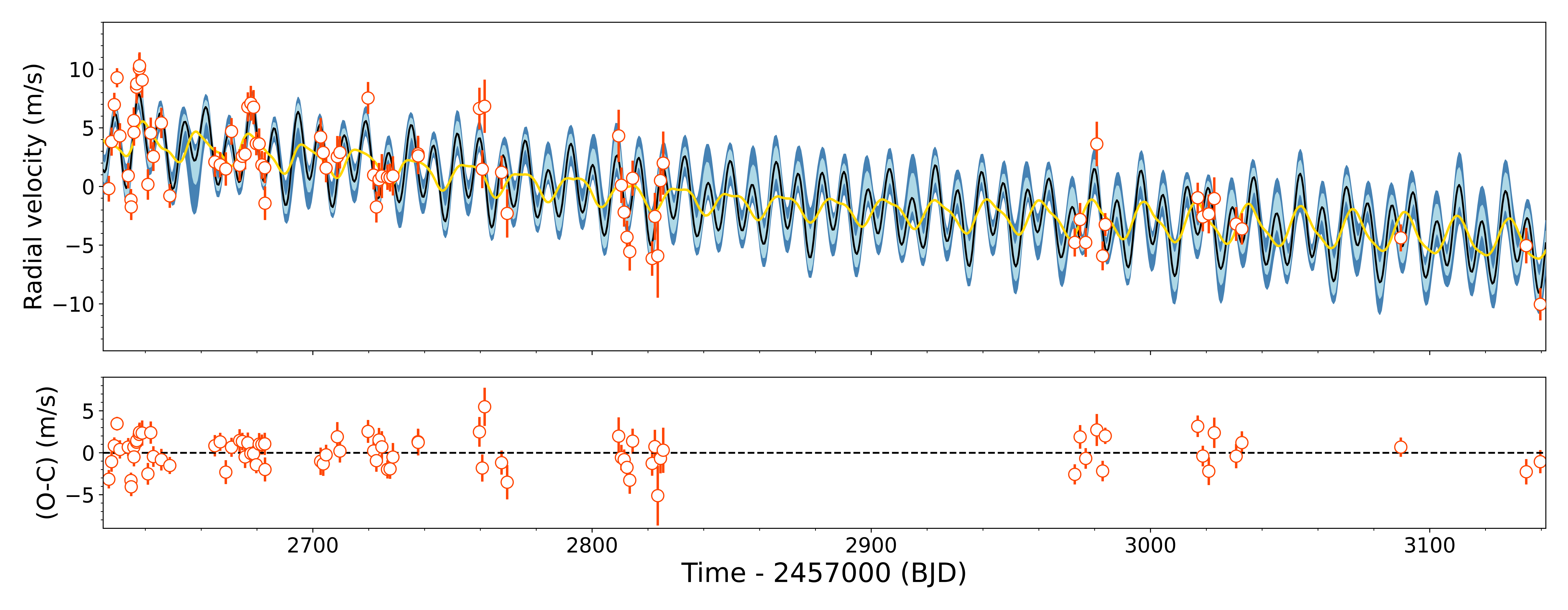}
    \includegraphics[width=\textwidth]{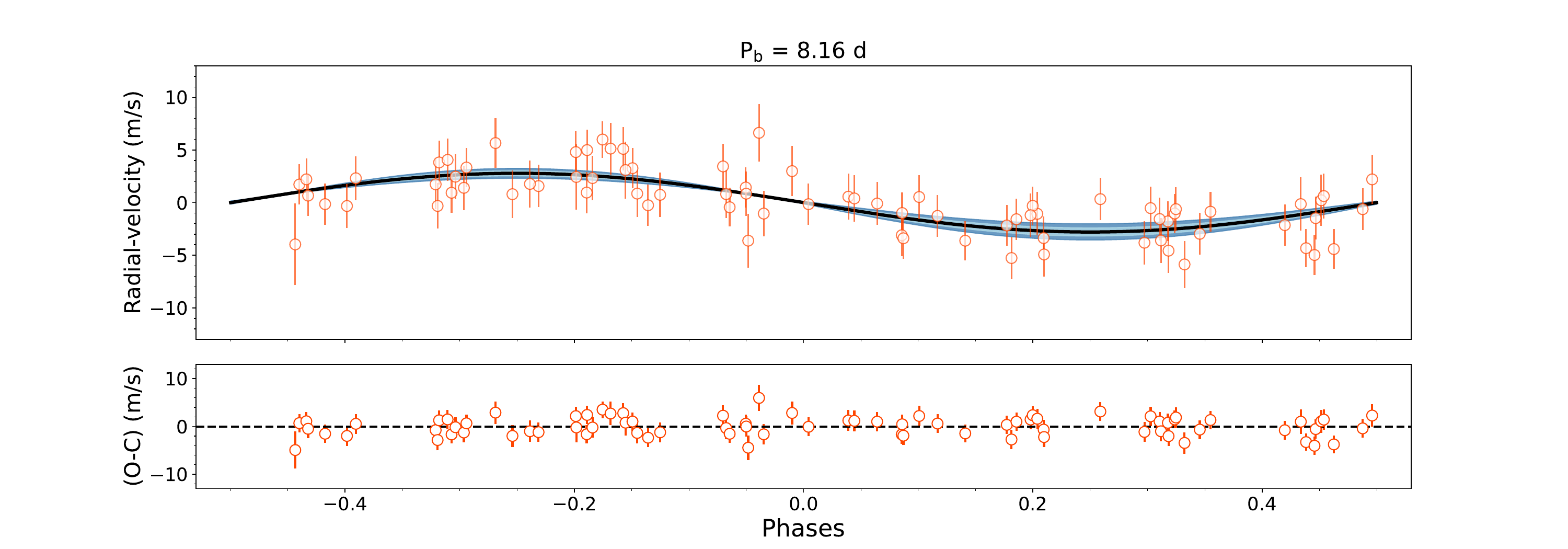}   
    \caption{Joint modeling of the RV data from HARPS (orange) for TOI-3493 along with their residuals.
    The solid curve is the median best-fit \texttt{juliet} model. 
    The shaded light and dark blue regions represent $68\,\%$ and $95\,\%$ credibility bands.
    {\em Top panel}:
    RV time-series with the GP component (solid yellow curve). 
    {\em Bottom panel}:
    Phase-folded RVs for TOI-3493 for the 8.15\,d-period planet along with the residuals folded at its period.}
    \label{fig:RV-joint-fit}
\end{figure*}

The posterior distribution of the planet parameters resulting from the joint orbital fit are summarized in Table \ref{tab:posteriors_planet} and Table \ref{tab:posteriors}. Furthermore, a covariance plot illustrating the fit parameters is presented in Fig. \ref{Fig:corner_plot-2pl}. Analysis of the posterior parameters obtained from our joint fit, along with the RV model depicted in Fig.~\ref{fig:RV-joint-fit}, reveals that the maximum posteriori of the GP periodic component, 
$P_\mathrm{rot,GP:RV}$, is $37.44^{+ 2.09}_{-1.07}$~d, with aliases observed at 34 and 41~d, as is seen in Fig.~\ref{fig:GP-Prot}. The best-fit results obtained from joint modeling are visualized in Figs. \ref{fig:photo-TESS} and \ref{fig:photo-TESS_phase} for transits and Fig.~\ref{fig:RV-joint-fit} for RVs.

    \begin{table}
    \centering
    \tiny
    \caption{Posterior parameters of the  
    joint fit for TOI-3493\,b}
    \label{tab:posteriors_planet}
    \begin{tabular}{lc} 
        \hline
        \hline
        \noalign{\smallskip}
        Parameter$^{a}$ & TOI-3493\,b \\
        \noalign{\smallskip}
        \hline
        \noalign{\smallskip}
        $P$ (d) & 
        $8.1594667^{+0.0000016}_{-0.0000017}$ \\[0.1 cm]
        $T_0$ (BJD) &      $2459313.0467455^{+0.0012477}_{-0.0012398}$  \\[0.1 cm]
        $a/R_\star$ &$14.07^{+0.31}_{-0.49}$ \\[0.1 cm]
        $b\,=\,(a/R_\star)\cos i_{\rm p}$  & $0.18^{+0.13}_{-0.12}$\\[0.1 cm]
        $i_{\rm p}$ (deg)   & $89.25^{+0.50}_{-0.59}$  \\[0.1 cm]
        $r_1$                                &$0.455^{+0.089}_{-0.081}$  \\[0.1 cm]
        $r_2$                                & $0.0240^{+0.0005}_{-0.0006}$  \\[0.1 cm]
        $K$ ($\mathrm{m\,s^{-1}}$)           & 
        $2.81^{+0.36}_{-0.37}$ \\[0.1 cm] 
        \noalign{\smallskip}
        \multicolumn{2}{l}{\it Derived physical parameters} \\
        \noalign{\smallskip}
        $M\,(M_\oplus)$       &$8.97^{+1.17}_{-1.20}$   \\[0.1 cm]  
        $R\,(R_\oplus)$       &$3.22^{+0.08}_{-0.08}$   \\[0.1 cm]
        $g\,(\rm{cm\,s^{-2}})$      &$8.47^{+1.19}_{-1.21}$   \\[0.1 cm]
        $\rho\, (\mathrm{g\,cm\,^{-3}}$) & $1.47^{+0.23}_{-0.22}$   \\[0.1 cm]
        $S\,({S_\oplus})$     &$245^{+18}_{-13}$   \\[0.1 cm]
       $T_{\rm eq}\,\rm(K)^b$    &$1102^{+20}_{-14}$   \\[0.1 cm]
    \noalign{\smallskip}
        \hline
    \end{tabular}
    \tablefoot{
      \tablefoottext{a}{Parameters obtained with the posterior values from Table\,\ref{tab:posteriors}.
      Error bars denote the 68\,\% posterior credibility intervals.}
      \tablefoottext{b}{The equilibrium temperature was calculated assuming zero Bond albedo.}
      }    
\end{table}

    \begin{table}
    \centering
    \tiny
    \caption{Posterior distributions of the \texttt{juliet} joint fit for the instrumental and GP fit parameters obtained for the TOI-3493 system.}
    \label{tab:posteriors}
    \begin{tabular}{lc} 
        \hline
        \hline
        \noalign{\smallskip}
        Parameter$^{a}$ & TOI-3493\\
        \noalign{\smallskip}
        \hline
        \noalign{\smallskip}    
        \multicolumn{2}{c}{\it Stellar parameters} \\[0.1cm]
        \noalign{\smallskip}
        $\rho_\star$ ($\mathrm{g\,cm\,^{-3}}$)& $0.790^{+0.053}_{-0.079}$\\[0.1 cm]
        \noalign{\smallskip}
        \multicolumn{2}{c}{\it Photometry parameters} \\[0.1cm]
        \noalign{\smallskip}
        $M_{\mathrm{TESS\,37}}$ ($10^{-6}$)   & $-21^{+4.19}_{-4.22}$\\[0.1 cm]
        $\sigma_{\mathrm{TESS\,37}}$ (ppm)     & $145^{+10.54}_{-11.27}$\\[0.1 cm]
        $q_{1,\mathrm{TESS\,37}}$               & $0.61^{+0.23}_{-0.22}$\\[0.1 cm]
        $q_{2,\mathrm{TESS\,37}}$               & $0.54^{+0.27}_{-0.26}$\\[0.1 cm]
        $M_{\mathrm{TESS\,64}}$ ($10^{-6}$)   & $-54^{+4.8}_{-4.7}$\\[0.1 cm]
        $\sigma_{\mathrm{TESS\,64}}$ (ppm)     & $287^{+6.7}_{-6.6}$\\[0.1 cm]
        $q_{1,\mathrm{TESS\,64}}$               & $0.31^{+0.28}_{-0.19}$\\[0.1 cm]
        $q_{2,\mathrm{TESS\,64}}$               & $0.29^{+0.23}_{-0.19}$\\[0.1 cm]
        $M_{\mathrm{LCO}}$ ($10^{-6}$)   & $115^{+54}_{-53}$\\[0.1 cm]
        $\sigma_{\mathrm{LCO}}$ (ppm)     & $901^{+45}_{-43}$\\[0.1 cm]
        $q_{1,\mathrm{LCO}}$               & $0.26^{+0.21}_{-0.19}$\\[0.1 cm]
        $q_{2,\mathrm{LCO}}$               & $0.42^{+0.35}_{-0.29}$\\[0.1 cm]
        \noalign{\smallskip}
        \multicolumn{2}{c}{\it RV parameters}\\[0.1cm]
        \noalign{\smallskip}
        $\mu_{\mathrm{HARPS}}$ ($\mathrm{m\,s^{-1}}$)       & $-2.29^{+4.9}_{-3.69}$\\[0.1 cm]
        $\sigma_{\mathrm{HARPS}}$ ($\mathrm{m\,s^{-1}}$)    & $1.6^{+0.32}_{-0.36}$\\[0.1cm]
        $\dot{\gamma}$ ($\mathrm{m\,s^{-1}day^{-1}}$)    & $-0.017\pm{0.002}$\\[0.1cm]
        \noalign{\smallskip}
        \multicolumn{2}{c}{\it GP hyperparameters} \\
        \noalign{\smallskip}
        $\sigma_\mathrm{GP,RV}$ ($\mathrm{m\,s^{-1}}$)              & $7.47^{+7.14}_{-3.38}$\\[0.1 cm]
        $\alpha_\mathrm{GP,RV}$ ($10^{-6}\,\mathrm{d^{-2}}$)        & $7.3^{+30.3}_{-6.74}$\\[0.1 cm]
        $\Gamma_\mathrm{GP,RV}$                           & $0.31^{+2.1}_{-0.28}$\\[0.1 cm]
        $P_\mathrm{rot;GP,RV}$ (d)                              & $37.44^{+ 2.09}_{-1.07}$\\[0.1 cm]
        \noalign{\smallskip}
        \hline
    \end{tabular}    \tablefoot{
      \tablefoottext{a}{Priors and descriptions for each parameter are in Table\,\ref{tab:priors}. Error bars denote the 68\,\% posterior credibility intervals.}}
    \end{table}

\section{Discussion} \label{sec:discussion}
\subsection{The mass-radius diagram}

\label{sec:mass_radius_discussion}

\begin{figure}[!ht]
\includegraphics[width=0.5\textwidth]{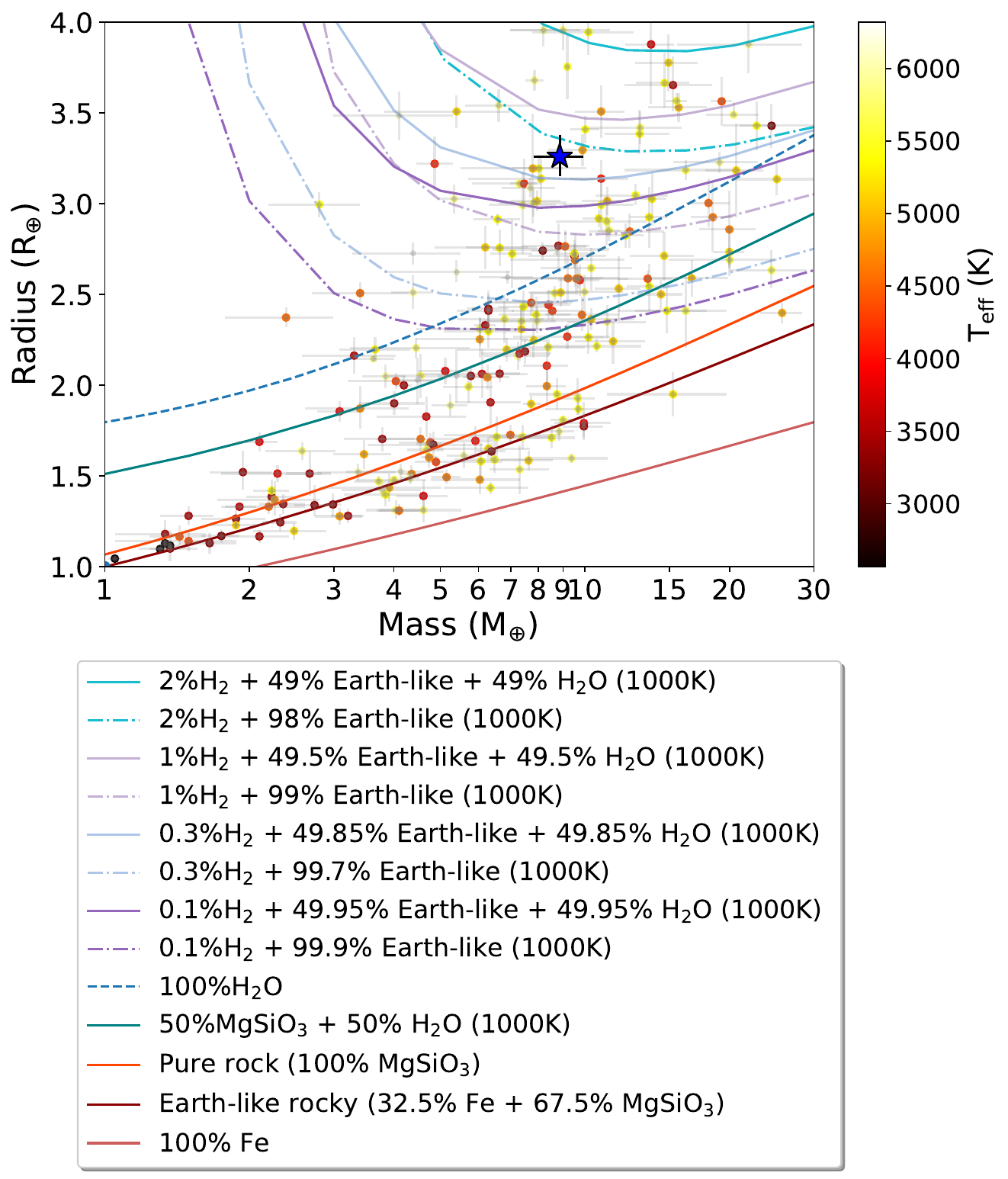}
\caption{Mass-radius diagram of well-characterized planets (that have masses and radii characterized with a precision better than $30\%$ and $10\%$, respectively) with radii of $R < 4\,R_{\oplus}$ taken from the Transiting Extrasolar Planet Catalogue \citep{Southworth2011}. The planets are color-coded based on the host stellar temperature, from hot (yellow) to cold (red). Theoretical M-R relations are taken from \cite{Zeng2019}. TOI-3493~b is marked with a blue star.}
\label{fig:mr_diagram}
\end{figure}

We derive the bulk density of TOI-3493 b as
$1.47^{+0.23}_{-0.22}~\mathrm{g\,cm^{-3}}$ based on the mass and radius obtained from our analysis (Sect.~\ref{subsubsec:joint}). Placing TOI-3493 b on the mass-radius (M-R) diagram, depicted in Fig.~\ref{fig:mr_diagram}, we compare it with other transiting planets sourced from the Exoplanet Archive\footnote{\url{https://exoplanetarchive.ipac.caltech.edu}.}. These planets are color-coded according to their host stellar temperatures. The solid and dashed lines represent models from \cite{Zeng2019}. We consider simplistic models using an H$_2$ envelope, and a core composed of H$_2$O and silicates of varying consistencies.

Two different models can best explain the position of TOI-3493~b on the M-R diagram:

1. Assuming a planetary equilibrium temperature of $\sim$1000~K (Table~\ref{tab:posteriors_planet}), we can consider the planetary core to consist of a $99.7\%$ Earth-like composition, with a rocky core of iron and silicates, and water in a 1:1 ratio, and the remaining $0.3\%$ comprising a hydrogen envelope. \\

2. Alternatively, the planetary core could be composed of $98\%$ iron and silicate, with the remaining $2\%$ constituting a hydrogen envelope.

This suggests that the planet may either have a large core predominantly composed of silicates and water with a thinner hydrogen outer layer, or a smaller, denser rocky core with a thicker hydrogen envelope.
In theory, many models can explain the position of TOI-3493~b with a larger percentage of water in the core or a larger fraction of H2 gas in the envelope. What can be concluded from these scenarios is that TOI-3493 b is likely a water-rich planet. Future atmospheric characterization studies can only shed more light on this case.

In recent years, the discovery of small planets with radii ranging from $1-4~R_{\oplus}$ has been abundant, with little similarities observed with our solar system planets. This distribution exhibits two distinct populations: one peaking around 
$1.5~R_{\oplus}$, termed super-Earths, and the other around $2.2~R_{\oplus}$, known as sub-Neptunes. The relative absence of planets between these peaks is referred to as the radius gap \citep{Fulton17, Vaneylen2018, Berger2018}.

Theories such as photo-evaporation \citep{Owen2013,Lopez2013} and core-powered mass loss \citep{Ginzburg+2018,Gupta2019} have been proposed to explain the radius gap, assuming similar initial planet compositions for both super-Earths and mini-Neptunes. While super-Earths lose their primordial atmospheres over time, leaving behind denser cores, mini-Neptunes retain their outer atmospheres. \cite{Leger2004} proposed an alternative model for mini-Neptune compositions, suggesting that their cores consist of an equal mixture of silicates and ice, known as ocean worlds or water worlds. This alternative composition can explain how planets with equally massive cores may have larger radii and lower densities, affecting their position on the M-R diagram \citep{Luque2022}. Consequently, models for the mini-Neptune population exhibit degeneracy. Studying planetary atmospheres, such as detecting hydrogen and helium in transmission spectroscopy or water molecules with JWST, could help resolve this degeneracy. Additionally, the radius gap may depend on the stellar age and evolve over millions of years to gigayears, particularly for models with hydrogen and helium in their outer envelopes, which cool over time due to their low molecular weight. In contrast, water worlds, composed of ice and silicates, may undergo less drastic evolution on similar timescales \citep{Rogers2023,Fernandes2022}.

\subsection{TOI-3493 b in the context of other exoplanets}

\begin{figure}[!ht]
\includegraphics[width=0.5\textwidth]{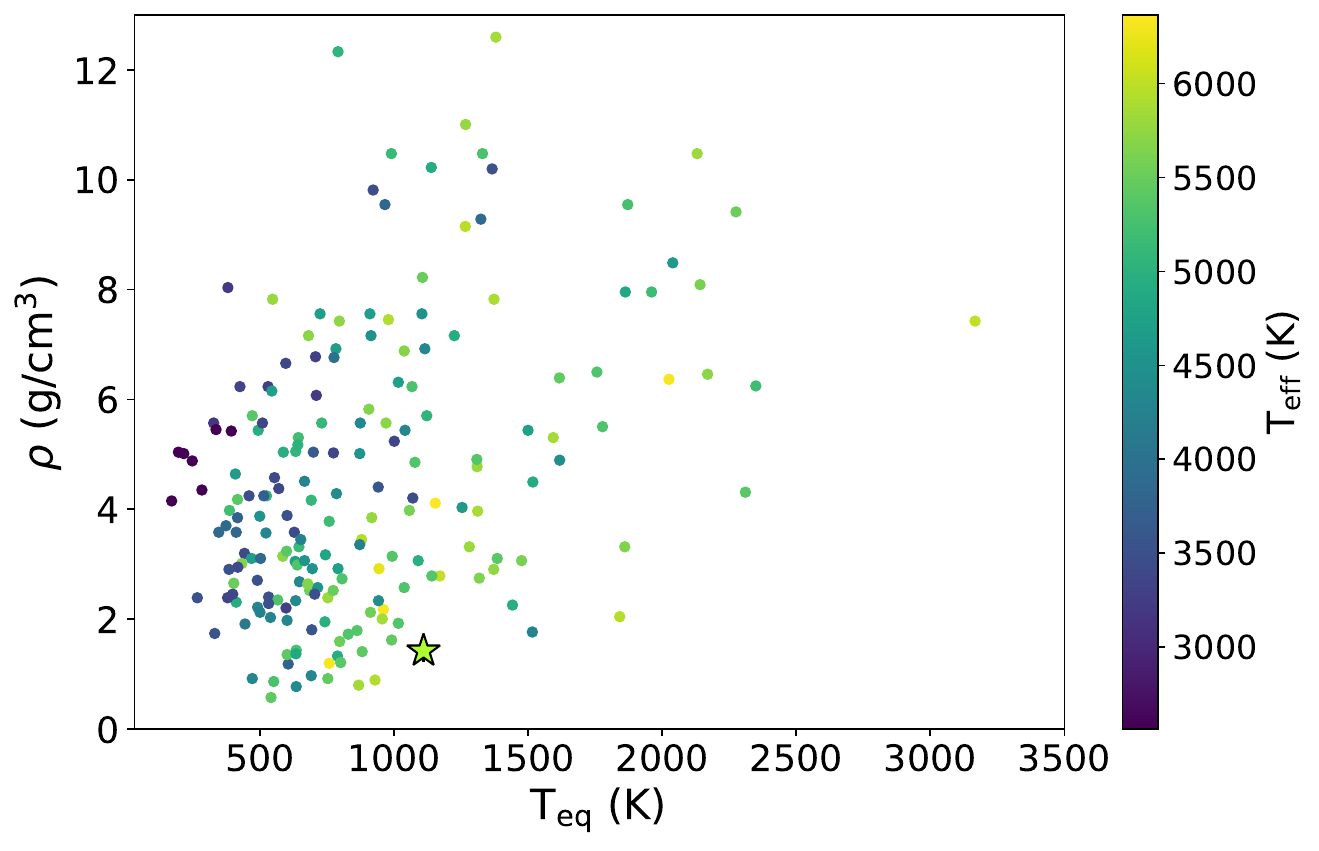}
\caption{Planet bulk density as a function of average equilibrium surface temperature of the planet taken from the exoplanet catalog (\citealt{Southworth2011}). The planets are color-coded based on the host stellar temperature, from hot (yellow) to cold (blue). The position of TOI-3493~b is marked with a star.}
\label{fig:rho_teq}
\end{figure}

In Fig.~\ref{fig:rho_teq}, we present a scatter plot illustrating the relationship between the bulk density ($\rho$) and the equilibrium surface temperature ($T_{\rm eq}$) of small planets with radii, $R_{\rm p}$, ranging from 1 to 4.0 $R_{\oplus}$. Notably, we include planets with actual mass measurements rather than solely relying on their $m \sin{i}$ values. The color scheme of the data points corresponds to the temperature of their host stars. We notice from the plot that $\rho$ increases with $T_{\rm eq}$, a consequence of the high irradiation at high $T_{\rm eq}$ evaporating away the less dense planets so that one ends up preferentially observing remnant cores or high mean molecular weight atmospheres.

The concentration of planets becomes denser from the top to the bottom of the plot. Specifically, the lower region, with planet densities roughly below 1 $\rho_{\oplus}$, predominantly comprises mini-Neptunes characterized by outer atmospheres, while the upper region encompasses super-Earths or terrestrial planets.

Solar-type stars exhibit a broad distribution of planets across all density ranges, whereas M dwarf stars tend to host a higher proportion of terrestrial planets compared to mini-Neptunes, which has also been confirmed in studies by \cite{Dressing2015,Mulders2015}. A distinctive fork-shaped pattern is evident in the distribution of terrestrial planets in Fig.~\ref{fig:rho_teq}. Planets orbiting M dwarfs are predominantly situated on the left side of the fork, reflecting their lower surface temperatures caused by reduced stellar insolation compared to those orbiting solar-type stars. Most stars on the right side of the fork have inflated radii, resulting in lower densities compared to M dwarf exoplanets with similar temperatures. TOI-3493~b, which orbits a G-type star, is located on the right side of the fork. Given its density, which is consistent with that of Neptune, it is hotter than any exoplanet of similar density around a cooler host star.
 The cluster of planets on the left side of the fork suggests a population exhibiting a weak dependence on equilibrium temperature relative to bulk density. The upper left grouping may consist of similar planets that have undergone atmospheric stripping, potentially due to XUV irradiation during their early evolution as was discussed previously in this paper, which is particularly relevant for small planets orbiting M dwarfs, which emit XUV radiation during their initial 100 Myrs. Analogous planets around solar-type stars are expected to have larger planetary cores or their outer envelopes scaled with respect to the stellar host mass, making them on average less dense than their M-dwarf counterparts for similar temperatures \citep{Petigura2022}. Thus, many of these are located on the right side of the fork.

\subsection{Prospects for atmospheric characterization}
\label{sec-patm}
Recently, \cite{Hord2024} presented a sample of \tess\  planets and planet candidates optimal for transmission and emission spectroscopy with \jwst{} in a given range of planetary sizes and equilibrium temperatures. All planets were divided into grids with bins in equilibrium temperature (\ppteq{}) from 100 to 3000 K ((100-350) K, (350-800) K, (800-1250) K, (1250-1750) K, (1750-2250) K, and (2250-3000) K) and planetary radius (\ppr{}) from 0.3 to 25.0 \Rea ((0.3-1.5) \Rea, (1.5-2.75) \Rea, (2.75-4.00)\Rea, (4-10) \Rea, and (10-25) \Rea) to identify the top planets and planet candidates ranked by the transmission spectroscopy metric (TSM) in each bin. This is justified by the fact that the TSM, as it is defined by \citet{Kempton2018}, Eq. (1), is proportional to the planet's equilibrium temperature and the cube of the planet's radius. Higher equilibrium temperatures and larger planetary radii simply increase the TSM very quickly.

\begin{figure}[!ht]
\centering
\includegraphics[width=0.925\linewidth]{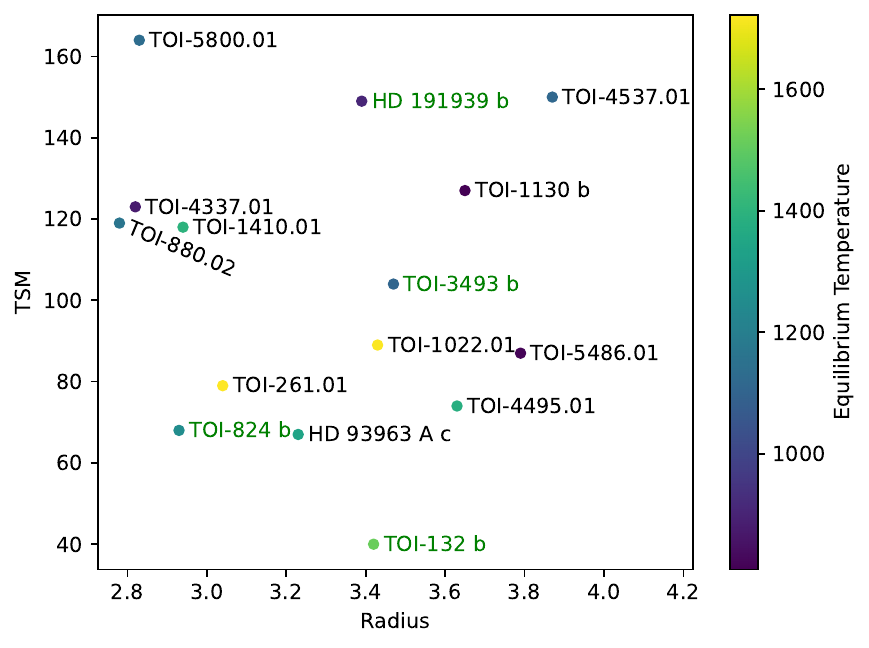}
    \caption{TSM versus planet radius for hot (800\,K < \ppteq{} < 1750\,K) large sub-Neptunes (2.75 < \ppr{} < 4.0\,\Rea). TOI-3493 b is  the second-most promising sub-Neptune along with three others (all marked in green) for transmission spectroscopy studies in the given temperature range and with mass measurement precisions exceeding $5\sigma$ discussed in the text. Data come from \cite{2024AJ....167..233H}. The names of the confirmed planets with mass measurements $> 5\sigma$ are in green.}
    \label{figure-TOI-3493-pr-tsm-800-1750-coloured_annotations}
\end{figure}

With a radius of \pbr = $3.22\pm{0.08}$\,\Rea{}, \planetb{} belongs to the group of large sub-Neptunes that \cite{2024AJ....167..233H} defined as planets with radii between 2.75 and 4.0\,\Rea. With \pbteq = $1102^{+20}_{-14}$\,\pptequ{}, \planetb{} is located close to the border of two ranges of equilibrium temperature defined by \cite{Hord2024} between 800\,K and 1250\,K and between 1250\,K and 1750\,K. Therefore, in Fig.~\ref{figure-TOI-3493-pr-tsm-800-1750-coloured_annotations} we present all \tess\  large sub-Neptunes and planetary candidates from the above two groups of equilibrium temperatures. As is shown in Fig.~\ref{figure-TOI-3493-pr-tsm-800-1750-coloured_annotations}, \planetb{} has a TSM of approximately 110, making it the second-most favorable target for transmission spectroscopy studies, among the four confirmed sub-Neptunes with mass measurement precisions exceeding $5\sigma$, highlighted with names in green font in the figure (with HD 191939 b being the most favorable among them).

\section{Summary} \label{sec:summary}
The TOI-3493 system comprises an early G-type star with a transiting sub-Neptune in a circular orbit. The host star has a surface temperature of $T_{\rm eff} = 5844\pm 42$\,K, surface gravity of $\log{g} = 4.25\pm0.06$\,dex, and metallicity of [Fe/H]\,=\,$0.03\pm0.04$\,dex. 
We thereby determine a stellar mass of $1.023\pm0.041\,M_{\odot}$ and a stellar radius of $1.228\pm0.017$\,$R_{\odot}$. 

The relatively bright star ($G$\,=\,9.127\,mag) is located at a distance of $96.76\pm0.2$\,pc. 
We also determine that the star is inactive, with a rotational period of around 34\,d.

The system consists of a transiting sub-Neptune that has a mass of $M$~=~$8.97^{+1.17}_{-1.20}$\,$M_\oplus$, a radius of $R = 3.22\pm{0.08}~R_{\oplus}$, and an orbital period of 8.15~d. The bulk density of the planet is $1.47^{+0.23}_{-0.22}$~g cm$^{-3}$, which is consistent with a Neptune-like density. TOI-3493 b is a promising candidate for transmission spectroscopy measurements for exoplanets, placed second to HD 191939 b on the metric scale. 

\section*{Data availability}
Table~B.1 is only available in electronic form at the CDS via anonymous ftp to cdsarc.u-strasbg.fr

\begin{acknowledgements}
PC  acknowledges the support of the Department of
Atomic Energy, Government of India and the financial support from the Th\"uringer Ministerium f\"ur Wirtschaft. The research is partly supported by Deutsche Forschungsgemeinschaft (DFG) grant CH 2636/1-1.
The spectroscopy observations were performed with the 3.6\,m telescope at the European Southern Observatory (La Silla, Chile) under programs 1102.C-0923 and 106.21TJ.001.
Funding for the \tess\  mission is provided by NASA's Science Mission Directorate. We acknowledge the use of public \tess\  data from pipelines at the \tess\  Science Office and at the \tess\  Science Processing Operations Center. This research has made use of the Exoplanet Follow-up Observation Program website, which is operated by the California Institute of Technology, under contract with the National Aeronautics and Space Administration under the Exoplanet Exploration Program. 
Resources supporting this work were provided by the NASA High-End Computing (HEC) Program through the NASA Advanced Supercomputing (NAS) Division at Ames Research Center for the production of the SPOC data products. This paper includes data collected by the \tess\  mission that are publicly available from the Mikulski Archive for Space Telescopes (MAST). 
This work makes use of observations from the LCOGT network. Part of the LCOGT telescope time was granted by NOIRLab through the Mid-Scale Innovations Program (MSIP). MSIP is funded by NSF. Funding for the \tess\  mission is provided by NASA's Science Mission Directorate. 

Some of the Observations in the paper made use of the High-Resolution Imaging instrument. ‘Alopeke. ‘Alopeke was funded by the NASA Exoplanet Exploration Program and built at the NASA Ames Research Center by Steve B. Howell, Nic Scott, Elliott P. Horch, and Emmett Quigley. Data were reduced using a software pipeline originally written by Elliott Horch and Mark Everett. ‘Alopeke was mounted on the Gemini North telescope of the international Gemini Observatory, a program of NSF’s OIR Lab, which is managed by the Association of Universities for Research in Astronomy (AURA) under a cooperative agreement with the National Science Foundation. on behalf of the Gemini partnership: the National Science Foundation (United States), National Research Council (Canada), Agencia Nacional de Investigación y Desarrollo (Chile),
Ministerio de Ciencia, Tecnolog\'ia e Innovaci\'on (Argentina), Minist\'erio da Ci\^encia, Tecnologia, Inova\c{c}\~oes e Comunica\c{c}\~oes (Brazil), and Korea Astronomy and Space Science Institute (Republic of Korea). 
Data were partly collected with the LCOGT network (part of the LCOGT telescope time was granted by
NOIRLab through the Mid-Scale Innovations Program (MSIP), which is funded
by the National Science Foundation). 
G.N. thanks for the research funding from the Ministry of Education and Science programme the "Excellence Initiative - Research University" conducted at the Centre of Excellence in Astrophysics and Astrochemistry of the Nicolaus Copernicus University in Toru\'n, Poland. G.N. also thanks the Centre of Informatics Tricity Academic Supercomputer and networK (CI TASK, Gda\'nsk, Poland) for providing computing resources (grant no. PT01187).

\end{acknowledgements}

\bibliographystyle{aa} 
\bibliography{bibliography}

\begin{appendix}

\onecolumn
\section{Priors and posteriors for data modeling}
\begin{table}[!h]
\small
    \centering
    \caption{Priors used for TOI-3493\,b in the  joint fit with \texttt{juliet}.}
    \label{tab:priors}
    \begin{tabular}{lccr} 
        \hline
        \hline
        \noalign{\smallskip}
        Parameter$^a$  & Prior & Unit & Description \\
        \noalign{\smallskip}
        \hline
        \noalign{\smallskip}
        \multicolumn{4}{c}{\it Stellar and planetary parameters} \\
        \noalign{\smallskip}
        $\rho_\star$ & $\mathcal{U}(100,10000)$ & g\,cm\,$^{-3}$ & Stellar density \\
        $P$              & $\mathcal{N}(8.159,0.002)$           & d                    & Period of planet b \\
        $t_{0}$                & $\mathcal{N}(2459313.0471,0.002)$     & d                    & Time of transit center of planet b \\
        $r_{1}$                & $\mathcal{U}(0,1)$                 & \dots                & Parameterization for $p$ and $b$ \\
        $r_{2}$                & $\mathcal{U}(0,1)$                 & \dots                & Parameterization for $p$ and $b$ \\
        $K$                  & $\mathcal{U}(0,20)$                & $\mathrm{m\,s^{-1}}$ & RV semi-amplitude of planet b \\
        $e$                  & 0.0 (fixed)                        & \dots                & Orbital eccentricity of planet b \\
        $\omega$             & 90.0 (fixed)                       & deg                  & Periastron angle of planet b \\
        \noalign{\smallskip} 
        \noalign{\smallskip}
        \multicolumn{4}{c}{\it Photometry parameters} \\
        $D_{\mathrm{TESS\,37,\,64}}$          & 1.0 (fixed)                  & \dots     & Dilution factor for \tess sectors 37, 64 \\
        $M_{\mathrm{TESS\,37,\,64}}$          & $\mathcal{N}(0,0.1)$         & \dots     & Relative flux offset for  \tess sectors 37, 64 \\
        $\sigma_{\mathrm{TESS\,37,\,64}}$     & $\mathcal{LU}(10^{-3},10^{4})$ & ppm       & Extra jitter term for sectors TESS 37, 64 \\
        $q_{1,\mathrm{TESS\,37,\,64}}$        & $\mathcal{U}(0,1)$           & \dots     & Limb-darkening parameterization for \tess sectors 37, 64\\
        $q_{2,\mathrm{TESS\,37,\,64}}$        & $\mathcal{U}(0,1)$           & \dots     & Limb-darkening parameterization for \tess sectors 37, 64\\
        \noalign{\smallskip}
        \multicolumn{4}{c}{\it RV parameters} \\
        \noalign{\smallskip}
        $\mu_{\mathrm{HARPS}}$            & $\mathcal{N}(-10,10)$  & $\mathrm{m\,s^{-1}}$ & RV zero point for HARPS \\
        $\sigma_{\mathrm{HARPS}}$         & $\mathcal{LU}(0.01,10)$ & $\mathrm{m\,s^{-1}}$ & Extra jitter term for HARPS \\
        $\dot{\gamma}$ & $\mathcal{U}(-100,100)$ & $\mathrm{m\,s^{-1}day^{-1}}$    & Linear trend/Radial acceleration\\
        \noalign{\smallskip}
        \noalign{\smallskip}
        \multicolumn{4}{c}{\it GP hyperparameters} \\
        \noalign{\smallskip}
        $\sigma_\mathrm{GP,RV}$  & $\mathcal{U}(0,100)$        & $\mathrm{m\,s^{-1}}$ & Amplitude of GP component for the RVs \\
        $\alpha_\mathrm{GP,RV}$  & $\mathcal{LU}(10^{-10},10^{-6})$ & d$^{-2}$             & Inverse length-scale of GP exponential component for the RVs\\
        $\Gamma_\mathrm{GP,RV}$  & $\mathcal{LU}(0.1,10)$       & \dots                & Amplitude of GP sine-squared component for the RVs \\
        $P_\mathrm{rot;GP,RV}$   & $\mathcal{U}(10,60)$        & d                    & Period of the GP QP component for the RVs \\
        \noalign{\smallskip}
        \hline
    \end{tabular}
    \tablefoot{
        \tablefoottext{a}{
$\mathcal{N}(\mu,\sigma^2)$ is a normal distribution of mean $\mu$ and variance $\sigma^2$, $\mathcal{U}(a,b)$ and $\mathcal{LU}(a,b)$ are uniform and log-uniform distributions between $a$ and $b$.}}
\end{table}

\begin{figure}[!h]
   \centering
   \includegraphics[width=0.5\textwidth]{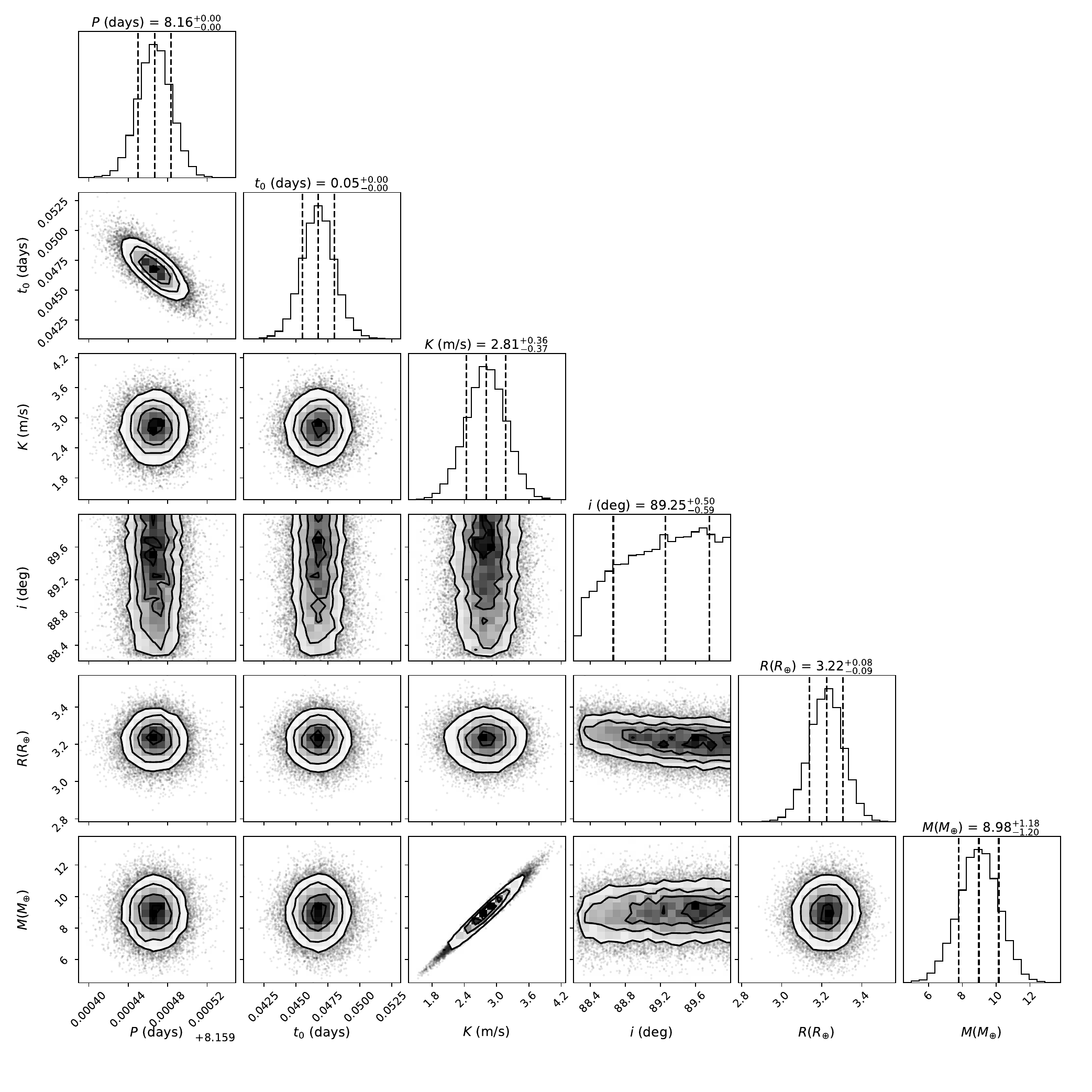}
   \caption{Posterior distribution for the joint model parameters (1cp+GP) derived with {\tt juliet}. The $t_{0}$ indicated on this plot is a truncated value ($t_{0}$ - 2459313) in BJD.}
    \label{Fig:corner_plot-2pl}
\end{figure}

\onecolumn
\section{Radial velocity table}\vspace{-1em}
\setlength{\tabcolsep}{4pt}
\small
\renewcommand{\arraystretch}{0.8} 
\begin{longtable}{l|c|c|c|c|c|c|c|c|c|c|}
\caption{
This table presents the HARPS radial velocity (RV) measurements and spectroscopic activity indicators for TOI-3493. Time stamps are in barycentric Julian day (BJD) in barycentric dynamical time (TDB). The RV measurements extracted using two pipelines (DRS and TERRA) are listed along with their respective uncertainties. For each RV data point, exposure time in seconds and signal to noise ratio (SNR) per pixel at 550 nm are given. The measurements of two profile diagnostics of the cross-correlation function, namely, the full-width-at-half-maximum (FWHM) and the bisector inverse slope (BIS), are provided along with four stellar activity indicators: Ca II H $\&$ K (log$R'_{HK}$),H$\alpha$, Na~{\sc i}~D$_1$ and D$_2$ lines.}\\
\hline
\noalign{\smallskip}
\multicolumn{1}{l|}{BJD$_{\rm TDB}$$^{a}$} & 
\multicolumn{1}{c|}{RV$_{\rm DRS}$} & 
\multicolumn{1}{c|}{RV$_{\rm TERRA}$} & 
\multicolumn{1}{c|}{Exposure} & 
\multicolumn{1}{c|}{SNR} & 
\multicolumn{1}{c|}{BIS} & 
\multicolumn{1}{c|}{FWHM} & 
\multicolumn{1}{c|}{log$R'_{HK}$} & 
\multicolumn{1}{c|}{H$\alpha$} & 
\multicolumn{1}{c|}{Na~{\sc i}~D$_1$} & 
\multicolumn{1}{c|}{Na~{\sc i}~D$_2$} \\
        \multicolumn{1}{l|}{} & 
        \multicolumn{1}{c|}{(km\,s$^{-1}$)} 
        & \multicolumn{1}{c|}{(km\,s$^{-1}$)} & \multicolumn{1}{c|}{(s)} & 
         \multicolumn{1}{c|}{(550nm)}&
         \multicolumn{1}{c|}{} & 
         \multicolumn{1}{c|}{}& 
         \multicolumn{1}{c|}{} & 
         \multicolumn{1}{c|}{}& 
         \multicolumn{1}{c|}{}&
         \multicolumn{1}{c|}{}  \\
\hline
\noalign{\smallskip}
\endfirsthead
\hline
\noalign{\smallskip}
\multicolumn{1}{c|}{BJD$_{\rm TDB}$$^{a}$} &
\multicolumn{1}{c|}{RV$_{DRS}$} 
& \multicolumn{1}{c|}{RV$_{TERRA}$} & \multicolumn{1}{c|}{Exposure} & \multicolumn{1}{c|}{SNR} & \multicolumn{1}{c|}{BIS} & \multicolumn{1}{c|}{FWHM} & \multicolumn{1}{c|}{log$R'_{HK}$} & \multicolumn{1}{c|}{H$\alpha$} & \multicolumn{1}{c|}{Na~{\sc i}~D$_1$} & \multicolumn{1}{c|}{Na~{\sc i}~D$_2$} \\
        \multicolumn{1}{l|}{} & \multicolumn{1}{c|}{(km\,s$^{-1}$)} & \multicolumn{1}{c|}{(km\,s$^{-1}$)} & \multicolumn{1}{c|}{(s)} &  \multicolumn{1}{c|}{(550nm)}  &\multicolumn{1}{c|}{} & \multicolumn{1}{c|}{}& \multicolumn{1}{c|}{} & \multicolumn{1}{c|}{}& \multicolumn{1}{c|}{} & \multicolumn{1}{c|}{}\\
\hline
\noalign{\smallskip}
\endhead
\hline
\endfoot
\setlength{\tabcolsep}{4pt}
2459626.880582 & $32.8110 \pm 0.0009$ & $-0.0017 \pm 0.0011$ & 1320 & 107 & 0.0107 & 7.1431 & $-4.9518 \pm 0.0089$ & 1.047 & 0.511 & 0.356 \\
2459627.860459	&	$	32.8166	\pm	0.0010	$	&	$	0.0024	\pm	0.0012	$	&	1320	&	102	&	0.0116	&	7.1508	&	$	-4.9214	\pm	0.0079	$	&	1.046	&	0.513	&	0.355	\\
2459628.866403	&	$	32.8174	\pm	0.0009	$	&	$	0.0055	\pm	0.0010	$	&	1320	&	108	&	0.0112	&	7.1473	&	$	-4.9427	\pm	0.0075	$	&	1.045	&	0.513	&	0.353	\\
2459629.835200	&	$	32.8194	\pm	0.0008	$	&	$	0.0078	\pm	0.0008	$	&	1320	&	125	&	0.0087	&	7.1482	&	$	-4.9471	\pm	0.0060	$	&	1.047	&	0.513	&	0.361	\\
2459630.854601	&	$	32.8166	\pm	0.0010	$	&	$	0.0028	\pm	0.0011	$	&	1200	&	97	&	0.0116	&	7.1552	&	$	-4.9230	\pm	0.0084	$	&	1.047	&	0.512	&	0.359	\\
2459633.858817	&	$	32.8115	\pm	0.0009	$	&	$	-0.0006	\pm	0.0011	$	&	1500	&	104	&	0.0130	&	7.1524	&	$	-4.9301	\pm	0.0078	$	&	1.049	&	0.513	&	0.355	\\
2459634.842599	&	$	32.8095	\pm	0.0009	$	&	$	-0.0027	\pm	0.0009	$	&	1200	&	110	&	0.0126	&	7.1445	&	$	-4.9408	\pm	0.0076	$	&	1.048	&	0.515	&	0.357	\\
2459634.902424	&	$	32.8082	\pm	0.0010	$	&	$	-0.0032	\pm	0.0011	$	&	1200	&	100	&	0.0111	&	7.1468	&	$	-4.9288	\pm	0.0084	$	&	1.046	&	0.510	&	0.359	\\
2459635.836321	&	$	32.8166	\pm	0.0010	$	&	$	0.0041	\pm	0.0011	$	&	1200	&	100	&	0.0153	&	7.1419	&	$	-4.9438	\pm	0.0092	$	&	1.045	&	0.513	&	0.359	\\
2459635.898924	&	$	32.8165	\pm	0.0010	$	&	$	0.0031	\pm	0.0011	$	&	1200	&	98	&	0.0117	&	7.1400	&	$	-4.9369	\pm	0.0084	$	&	1.046	&	0.513	&	0.357	\\
2459636.831425	&	$	32.8187	\pm	0.0012	$	&	$	0.0070	\pm	0.0014	$	&	1200	&	79	&	0.0112	&	7.1385	&	$	-4.9280	\pm	0.0113	$	&	1.045	&	0.511	&	0.355	\\
2459636.891874	&	$	32.8204	\pm	0.0010	$	&	$	0.0073	\pm	0.0013	$	&	1200	&	99	&	0.0122	&	7.1414	&	$	-4.9606	\pm	0.0090	$	&	1.042	&	0.513	&	0.355	\\
2459637.804621	&	$	32.8212	\pm	0.0010	$	&	$	0.0086	\pm	0.0013	$	&	1200	&	96	&	0.0133	&	7.1461	&	$	-4.9113	\pm	0.0076	$	&	1.043	&	0.507	&	0.356	\\
2459637.884848	&	$	32.8206	\pm	0.0010	$	&	$	0.0088	\pm	0.0011	$	&	1200	&	101	&	0.0140	&	7.1463	&	$	-4.9329	\pm	0.0083	$	&	1.043	&	0.509	&	0.356	\\
2459638.852126	&	$	32.8199	\pm	0.0011	$	&	$	0.0075	\pm	0.0015	$	&	1200	&	89	&	0.0164	&	7.1460	&	$	-4.9413	\pm	0.0099	$	&	1.041	&	0.512	&	0.358	\\
2459640.905386	&	$	32.8104	\pm	0.0013	$	&	$	-0.0013	\pm	0.0013	$	&	1200	&	74	&	0.0134	&	7.1421	&	$	-4.9539	\pm	0.0122	$	&	1.042	&	0.505	&	0.350	\\
2459641.895396	&	$	32.8151	\pm	0.0010	$	&	$	0.0031	\pm	0.0013	$	&	1200	&	97	&	0.0129	&	7.1453	&	$	-4.9386	\pm	0.0089	$	&	1.045	&	0.511	&	0.353	\\
2459642.850524	&	$	32.8123	\pm	0.0010	$	&	$	0.0011	\pm	0.0012	$	&	1200	&	95	&	0.0109	&	7.1414	&	$	-4.9538	\pm	0.0094	$	&	1.045	&	0.514	&	0.353	\\
2459645.698590	&	$	32.8159	\pm	0.0011	$	&	$	0.0039	\pm	0.0013	$	&	1200	&	86	&	0.0140	&	7.1435	&	$	-5.0035	\pm	0.0101	$	&	1.047	&	0.512	&	0.349	\\
2459648.735477	&	$	32.8092	\pm	0.0008	$	&	$	-0.0023	\pm	0.0010	$	&	1200	&	111	&	0.0111	&	7.1436	&	$	-4.9783	\pm	0.0066	$	&	1.047	&	0.511	&	0.355	\\
2459664.857124	&	$	32.8120	\pm	0.0011	$	&	$	0.0006	\pm	0.0013	$	&	1200	&	92	&	0.0148	&	7.1373	&	$	-5.0086	\pm	0.0109	$	&	1.051	&	0.516	&	0.355	\\
2459666.799913	&	$	32.8133	\pm	0.0010	$	&	$	0.0004	\pm	0.0011	$	&	1200	&	92	&	0.0094	&	7.1447	&	$	-4.9473	\pm	0.0089	$	&	1.052	&	0.511	&	0.353	\\
2459668.813333	&	$	32.8148	\pm	0.0014	$	&	$	0.0000	\pm	0.0014	$	&	1200	&	69	&	0.0116	&	7.1433	&	$	-4.9536	\pm	0.0125	$	&	1.053	&	0.512	&	0.360	\\
2459670.845583	&	$	32.8142	\pm	0.0010	$	&	$	0.0032	\pm	0.0011	$	&	1200	&	99	&	0.0150	&	7.1469	&	$	-4.9587	\pm	0.0091	$	&	1.050	&	0.511	&	0.358	\\
2459673.726708	&	$	32.8139	\pm	0.0012	$	&	$	0.0003	\pm	0.0014	$	&	1200	&	80	&	0.0113	&	7.1419	&	$	-4.9567	\pm	0.0107	$	&	1.054	&	0.509	&	0.359	\\
2459674.704683	&	$	32.8134	\pm	0.0009	$	&	$	0.0011	\pm	0.0010	$	&	1200	&	106	&	0.0081	&	7.1411	&	$	-4.9573	\pm	0.0069	$	&	1.052	&	0.514	&	0.354	\\
2459675.708157	&	$	32.8139	\pm	0.0009	$	&	$	0.0013	\pm	0.0013	$	&	1200	&	104	&	0.0115	&	7.1395	&	$	-4.9426	\pm	0.0071	$	&	1.053	&	0.517	&	0.354	\\
2459676.683585	&	$	32.8183	\pm	0.0011	$	&	$	0.0053	\pm	0.0012	$	&	1200	&	84	&	0.0145	&	7.1399	&	$	-4.9595	\pm	0.0087	$	&	1.050	&	0.515	&	0.356	\\
2459677.745373	&	$	32.8178	\pm	0.0010	$	&	$	0.0056	\pm	0.0015	$	&	1200	&	90	&	0.0085	&	7.1419	&	$	-4.9510	\pm	0.0080	$	&	1.049	&	0.509	&	0.356	\\
2459678.720420	&	$	32.8168	\pm	0.0011	$	&	$	0.0053	\pm	0.0015	$	&	1200	&	86	&	0.0070	&	7.1497	&	$	-4.9637	\pm	0.0087	$	&	1.050	&	0.514	&	0.351	\\
2459679.696709	&	$	32.8153	\pm	0.0010	$	&	$	0.0022	\pm	0.0010	$	&	1200	&	91	&	0.0107	&	7.1380	&	$	-4.9416	\pm	0.0078	$	&	1.049	&	0.511	&	0.358	\\
2459680.746406	&	$	32.8145	\pm	0.0012	$	&	$	0.0021	\pm	0.0013	$	&	1200	&	82	&	0.0081	&	7.1418	&	$	-4.9420	\pm	0.0101	$	&	1.050	&	0.511	&	0.358	\\
2459681.737777	&	$	32.8122	\pm	0.0010	$	&	$	0.0003	\pm	0.0012	$	&	1200	&	92	&	0.0070	&	7.1415	&	$	-4.9616	\pm	0.0093	$	&	1.051	&	0.514	&	0.353	\\
2459682.756872	&	$	32.8113	\pm	0.0011	$	&	$	0.0001	\pm	0.0013	$	&	1200	&	87	&	0.0116	&	7.1356	&	$	-4.9494	\pm	0.0103	$	&	1.051	&	0.516	&	0.356	\\
2459682.820267	&	$	32.8094	\pm	0.0011	$	&	$	-0.0029	\pm	0.0014	$	&	1200	&	91	&	0.0100	&	7.1412	&	$	-4.9651	\pm	0.0103	$	&	1.051	&	0.517	&	0.351	\\
2459702.753589	&	$	32.8155	\pm	0.0014	$	&	$	0.0028	\pm	0.0016	$	&	1200	&	68	&	0.0068	&	7.1435	&	$	-4.9234	\pm	0.0132	$	&	1.048	&	0.514	&	0.357	\\
2459703.680296	&	$	32.8152	\pm	0.0012	$	&	$	0.0014	\pm	0.0014	$	&	1200	&	82	&	0.0130	&	7.1386	&	$	-4.9700	\pm	0.0109	$	&	1.051	&	0.514	&	0.352	\\
2459704.736398	&	$	32.8132	\pm	0.0011	$	&	$	0.0000	\pm	0.0012	$	&	1200	&	84	&	0.0090	&	7.1414	&	$	-4.9545	\pm	0.0097	$	&	1.050	&	0.512	&	0.354	\\
2459708.747806	&	$	32.8167	\pm	0.0013	$	&	$	0.0010	\pm	0.0018	$	&	1200	&	73	&	0.0085	&	7.1407	&	$	-4.9599	\pm	0.0133	$	&	1.050	&	0.521	&	0.352	\\
2459709.673948	&	$	32.8156	\pm	0.0011	$	&	$	0.0014	\pm	0.0014	$	&	1200	&	89	&	0.0133	&	7.1417	&	$	-4.9428	\pm	0.0091	$	&	1.048	&	0.519	&	0.358	\\
2459719.736469	&	$	32.8182	\pm	0.0013	$	&	$	0.0061	\pm	0.0014	$	&	1200	&	76	&	0.0028	&	7.1489	&	$	-4.9434	\pm	0.0125	$	&	1.050	&	0.526	&	0.350	\\
2459721.721612	&	$	32.8144	\pm	0.0013	$	&	$	-0.0005	\pm	0.0012	$	&	1200	&	76	&	0.0079	&	7.1398	&	$	-4.9903	\pm	0.0131	$	&	1.045	&	0.518	&	0.350	\\
2459722.728635	&	$	32.8102	\pm	0.0011	$	&	$	-0.0033	\pm	0.0013	$	&	1200	&	86	&	0.0037	&	7.1504	&	$	-4.9738	\pm	0.0123	$	&	1.046	&	0.525	&	0.348	\\
2459723.670279	&	$	32.8117	\pm	0.0012	$	&	$	-0.0009	\pm	0.0014	$	&	1800	&	78	&	0.0113	&	7.1464	&	$	-4.9705	\pm	0.0114	$	&	1.044	&	0.515	&	0.354	\\
2459724.703948	&	$	32.8116	\pm	0.0018	$	&	$	-0.0006	\pm	0.0019	$	&	1200	&	54	&	0.0095	&	7.1448	&	$	-4.9597	\pm	0.0190	$	&	1.049	&	0.512	&	0.359	\\
2459726.674096	&	$	32.8131	\pm	0.0009	$	&	$	-0.0007	\pm	0.0011	$	&	1200	&	103	&	0.0052	&	7.1448	&	$	-4.9611	\pm	0.0078	$	&	1.051	&	0.517	&	0.360	\\
2459727.635942	&	$	32.8114	\pm	0.0010	$	&	$	-0.0007	\pm	0.0012	$	&	1200	&	91	&	0.0023	&	7.1443	&	$	-4.9640	\pm	0.0093	$	&	1.049	&	0.511	&	0.360	\\
2459728.628279	&	$	32.8122	\pm	0.0013	$	&	$	-0.0006	\pm	0.0017	$	&	1200	&	74	&	0.0116	&	7.1443	&	$	-4.9451	\pm	0.0125	$	&	1.048	&	0.520	&	0.363	\\
2459737.657637	&	$	32.8144	\pm	0.0014	$	&	$	0.0013	\pm	0.0015	$	&	1200	&	70	&	0.0089	&	7.1520	&	$	-4.9620	\pm	0.0136	$	&	1.048	&	0.519	&	0.359	\\
2459737.700305	&	$	32.8125	\pm	0.0014	$	&	$	0.0011	\pm	0.0016	$	&	1200	&	67	&	0.0072	&	7.1451	&	$	-4.9382	\pm	0.0145	$	&	1.049	&	0.517	&	0.357	\\
2459759.624560	&	$	32.8169	\pm	0.0015	$	&	$	0.0051	\pm	0.0018	$	&	1200	&	66	&	0.0103	&	7.1504	&	$	-4.9523	\pm	0.0147	$	&	1.046	&	0.507	&	0.356	\\
2459760.633238	&	$	32.8137	\pm	0.0014	$	&	$	0.0000	\pm	0.0016	$	&	1200	&	70	&	0.0109	&	7.1480	&	$	-4.9200	\pm	0.0118	$	&	1.048	&	0.516	&	0.361	\\
2459761.500397	&	$	32.8161	\pm	0.0020	$	&	$	0.0053	\pm	0.0023	$	&	1200	&	48	&	0.0067	&	7.1392	&	$	-4.9643	\pm	0.0191	$	&	1.042	&	0.507	&	0.362	\\
2459767.559214	&	$	32.8158	\pm	0.0012	$	&	$	-0.0003	\pm	0.0014	$	&	1200	&	79	&	0.0124	&	7.1428	&	$	-4.9312	\pm	0.0110	$	&	1.046	&	0.510	&	0.362	\\
2459769.582880	&	$	32.8099	\pm	0.0018	$	&	$	-0.0039	\pm	0.0021	$	&	1200	&	53	&	0.0025	&	7.1438	&	$	-5.0147	\pm	0.0224	$	&	1.049	&	0.507	&	0.368	\\
2459809.507084	&	$	32.8203	\pm	0.0019	$	&	$	0.0028	\pm	0.0022	$	&	1800	&	55	&	0.0056	&	7.1454	&	$	-4.9828	\pm	0.0222	$	&	1.044	&	0.505	&	0.366	\\
2459810.490819	&	$	32.8152	\pm	0.0012	$	&	$	-0.0014	\pm	0.0015	$	&	1200	&	85	&	0.0089	&	7.1527	&	$	-4.9917	\pm	0.0131	$	&	1.050	&	0.512	&	0.362	\\
2459811.473077	&	$	32.8117	\pm	0.0012	$	&	$	-0.0037	\pm	0.0012	$	&	1200	&	85	&	0.0056	&	7.1577	&	$	-4.9822	\pm	0.0118	$	&	1.050	&	0.510	&	0.362	\\
2459812.485712	&	$	32.8086	\pm	0.0012	$	&	$	-0.0058	\pm	0.0014	$	&	1200	&	80	&	0.0098	&	7.1521	&	$	-5.0047	\pm	0.0130	$	&	1.049	&	0.512	&	0.365	\\
2459813.485623	&	$	32.8067	\pm	0.0014	$	&	$	-0.0070	\pm	0.0016	$	&	1200	&	71	&	0.0131	&	7.1612	&	$	-5.0278	\pm	0.0172	$	&	1.049	&	0.510	&	0.363	\\
2459814.476853	&	$	32.8151	\pm	0.0013	$	&	$	-0.0008	\pm	0.0015	$	&	1500	&	75	&	0.0097	&	7.1486	&	$	-4.9627	\pm	0.0127	$	&	1.042	&	0.514	&	0.358	\\
2459821.477328	&	$	32.8050	\pm	0.0011	$	&	$	-0.0076	\pm	0.0015	$	&	1500	&	88	&	0.0092	&	7.1560	&	$	-4.9995	\pm	0.0129	$	&	1.045	&	0.513	&	0.361	\\
2459822.472553	&	$	32.8072	\pm	0.0018	$	&	$	-0.0041	\pm	0.0020	$	&	1200	&	56	&	0.0017	&	7.1612	&	$	-5.0608	\pm	0.0276	$	&	1.040	&	0.505	&	0.360	\\
2459823.475169	&	$	32.8044	\pm	0.0035	$	&	$	-0.0074	\pm	0.0036	$	&	1500	&	32	&	-0.0014	&	7.1562	&	$	-5.1361	\pm	0.0802	$	&	1.047	&	0.510	&	0.353	\\
2459824.474586	&	$	32.8117	\pm	0.0016	$	&	$	-0.0010	\pm	0.0018	$	&	1500	&	63	&	0.0109	&	7.1571	&	$	-5.0002	\pm	0.0234	$	&	1.049	&	0.517	&	0.360	\\
2459825.475658	&	$	32.8132	\pm	0.0026	$	&	$	0.0005	\pm	0.0027	$	&	1500	&	41	&	0.0141	&	7.1604	&	$	-5.0723	\pm	0.0488	$	&	1.044	&	0.522	&	0.359	\\
2459972.856727	&	$	32.8061	\pm	0.0009	$	&	$	-0.0063	\pm	0.0012	$	&	1200	&	103	&	0.0112	&	7.1525	&	$	-4.9670	\pm	0.0076	$	&	1.049	&	0.513	&	0.357	\\
2459974.786337	&	$	32.8076	\pm	0.0011	$	&	$	-0.0043	\pm	0.0014	$	&	1200	&	85	&	0.0103	&	7.1486	&	$	-4.9594	\pm	0.0099	$	&	1.038	&	0.511	&	0.358	\\
2459976.783459	&	$	32.8079	\pm	0.0010	$	&	$	-0.0062	\pm	0.0012	$	&	1200	&	97	&	0.0110	&	7.1525	&	$	-4.9679	\pm	0.0084	$	&	1.041	&	0.504	&	0.341	\\
2459980.749082	&	$	32.8149	\pm	0.0018	$	&	$	0.0022	\pm	0.0019	$	&	1200	&	57	&	0.0047	&	7.1675	&	$	-4.8728	\pm	0.0164	$	&	1.040	&	0.513	&	0.341	\\
2459982.833378	&	$	32.8091	\pm	0.0010	$	&	$	-0.0074	\pm	0.0012	$	&	1200	&	98	&	0.0183	&	7.1541	&	$	-4.9757	\pm	0.0084	$	&	1.047	&	0.513	&	0.348	\\
2459983.753814	&	$	32.8077	\pm	0.0011	$	&	$	-0.0047	\pm	0.0010	$	&	1200	&	87	&	0.0167	&	7.1529	&	$	-4.9915	\pm	0.0111	$	&	1.044	&	0.511	&	0.341	\\
2460016.873133	&	$	32.8108	\pm	0.0011	$	&	$	-0.0024	\pm	0.0013	$	&	1200	&	91	&	0.0163	&	7.1516	&	$	-4.9875	\pm	0.0110	$	&	1.047	&	0.515	&	0.345	\\
2460018.740283	&	$	32.8091	\pm	0.0011	$	&	$	-0.0041	\pm	0.0012	$	&	1200	&	90	&	0.0130	&	7.1560	&	$	-4.9975	\pm	0.0098	$	&	1.049	&	0.519	&	0.345	\\
2460020.848260	&	$	32.8084	\pm	0.0013	$	&	$	-0.0038	\pm	0.0017	$	&	1200	&	78	&	0.0097	&	7.1513	&	$	-4.9775	\pm	0.0130	$	&	1.052	&	0.512	&	0.345	\\
2460022.838660	&	$	32.8108	\pm	0.0018	$	&	$	-0.0025	\pm	0.0018	$	&	1200	&	57	&	0.0126	&	7.1576	&	$	-5.0789	\pm	0.0271	$	&	1.051	&	0.518	&	0.349	\\
2460030.671688	&	$	32.8072	\pm	0.0012	$	&	$	-0.0047	\pm	0.0014	$	&	1200	&	76	&	0.0136	&	7.1454	&	$	-4.9886	\pm	0.0106	$	&	1.053	&	0.519	&	0.344	\\
2460032.695311	&	$	32.8067	\pm	0.0011	$	&	$	-0.0051	\pm	0.0013	$	&	1200	&	86	&	0.0153	&	7.1517	&	$	-4.9825	\pm	0.0100	$	&	1.055	&	0.515	&	0.348	\\
2460089.642258	&	$	32.8082	\pm	0.0010	$	&	$	-0.0059	\pm	0.0011	$	&	1200	&	95	&	0.0096	&	7.1484	&	$	-4.9831	\pm	0.0089	$	&	1.048	&	0.519	&	0.359	\\
2460134.548130	&	$	32.8085	\pm	0.0016	$	&	$	-0.0066	\pm	0.0015	$	&	1200	&	62	&	0.0121	&	7.1530	&	$	-4.9735	\pm	0.0181	$	&	1.053	&	0.513	&	0.362	\\
2460139.579002	&	$	32.8025	\pm	0.0014	$	&	$	-0.0115	\pm	0.0014	$	&	1200	&	71	&	0.0031	&	7.1387	&	$	-4.9652	\pm	0.0141	$	&	1.050	&	0.514	&	0.362	\\
2460173.525400	&	$	32.8021	\pm	0.0021	$	&	$	-0.0115	\pm	0.0019	$	&	1200	&	48	&	0.0110	&	7.1386	&	$	-4.9942	\pm	0.0298	$	&	1.052	&	0.510	&	0.356	\\
\label{tab:RV_table}
\end{longtable}

\tablefoot{
(a) Barycentric Julian date in the barycentric dynamical time standard. 
(b) When not mentioned, error bars are of the order of $10\%$ on individual values in the table.
}

\end{appendix}

\end{document}